\documentclass[useAMS,usenatbib,usegraphicx]{mn2e}

\usepackage{graphicx}
\usepackage{natbib}
\usepackage{amsmath}
\usepackage{mathtools}
\usepackage{amssymb}
\usepackage{wasysym}
\usepackage{txfonts}
\usepackage[usenames, dvipsnames]{color}

\usepackage{latexsym,longtable,epsf,ulem}
\usepackage{cases} 
\usepackage{enumerate} 

\newcommand{\ant}{\alpha_{\rm nt}}
\newcommand{\fth}{f_{\rm th}}

\newcommand{\rad}{rad m$^{-2}$~}
\newcommand{\sigmaRM}{\sigma_{\rm RM}}

\title[The Antennae galaxies]
{Detection of a $\sim$20 kpc coherent magnetic field in the outskirt of merging spirals: the Antennae galaxies}
\author[Basu et al.]{Aritra Basu$^1$\thanks{E-mail: abasu@mpifr-bonn.mpg.de (AB); mao@mpifr-bonn.mpg.de (SAM)}, 
	S. A. Mao$^{1\star}$, Amanda A. Kepley$^2$, Timothy Robishaw$^3$, Ellen G. Zweibel$^4$,
        \newauthor John. S. Gallagher III$^4$\\
$^1$Max-Planck-Institut f{\"u}r Radioastronomie, Auf dem H{\"u}gel 69, D-53121 Bonn, Germany\\
$^2$National Radio Astronomy Observatory, 520 Edgemont Road, Charlottesville, VA 22903-2475, USA \\
$^3$National Research Council Canada, Herzberg Astronomy and Astrophysics Programs, Dominion Radio Astrophysical Observatory, Penticton, BC V2A 6J9, Canada\\
$^4$Department of Astronomy and Physics, University of Wisconsin-Madison, WI 53706, USA}

\begin{document}


\pagerange{\pageref{firstpage}--\pageref{lastpage}} \pubyear{2002}

\maketitle

\label{firstpage}

\begin{abstract}

We present a study of the magnetic field properties of NGC~4038/9 (the
`Antennae' galaxies), the closest example of a late stage merger of two spiral
galaxies. Wideband polarimetric observations were performed using the Karl
G. Jansky Very Large Array between 2 and 4 GHz.  Rotation measure synthesis and
Faraday depolarization analysis was performed to probe the magnetic field
strength and structure at spatial resolution of $\sim1$ kpc. Highly polarized
emission from the southern tidal tail is detected with intrinsic fractional
polarization close to the theoretical maximum ($0.62\pm0.18$), estimated by
fitting the Faraday depolarization with a volume that is both synchrotron
emitting and Faraday rotating containing random magnetic fields.  Magnetic
fields are well aligned along the tidal tail and the Faraday depths shows
large-scale smooth variations preserving its sign.  This suggests the field in
the plane of the sky to be regular up to $\sim20$ kpc, which is the largest
detected regular field structure on galactic scales.  The equipartition
field strength of $\sim8.5~\mu$G of the regular field in the tidal tail is
reached within a few 100 Myr, likely generated by stretching of the galactic
disc field by a factor of 4--9 during the tidal interaction.  The regular field
strength is greater than the turbulent fields in the tidal tail.  Our study
comprehensively demonstrates, although the magnetic fields within the merging
bodies are dominated by strong turbulent magnetic fields of $\sim20~\mu$G in
strength, tidal interactions can produce large-scale regular field structure in
the outskirts.

\end{abstract}


\section{Introduction}

Magnetic fields are pervasive in the Universe on all scales and they play
crucial roles in various processes in the interstellar medium. The large-scale
ordered magnetic fields ($\gtrsim1$ kpc) in galaxies are thought to be
amplified via the $\alpha$--$\Omega$ dynamo mechanism---the buildup of
initially weak seed fields ($<10^{-9}$ G; \citealt{ade15}) to microgauss fields
via small-scale turbulence and differential rotation \citep{ruzma88, kulsr08}.
The small-scale dynamo can efficiently amplify the magnetic fields on scales
$\lesssim1$ kpc in $\sim10^6$ years \citep[][]{kandu99, feder11, chama13,
schob13}. On the other hand, the conventional $\alpha$--$\Omega$ dynamo action
requires $\sim10^9$ years to amplify the large-scale magnetic field in galaxies
\citep{arsha09, pakmo14}, which is too long to explain the detection of
coherent fields in young systems \citep[e.g.,][]{berne08, farne14}. This
suggests that there must be other magnetic field amplification processes at
work.

\begin{figure*}
\begin{tabular}{cc}
{\mbox{\includegraphics[width=9.2cm, trim=5mm 10mm 2mm 0mm,clip]{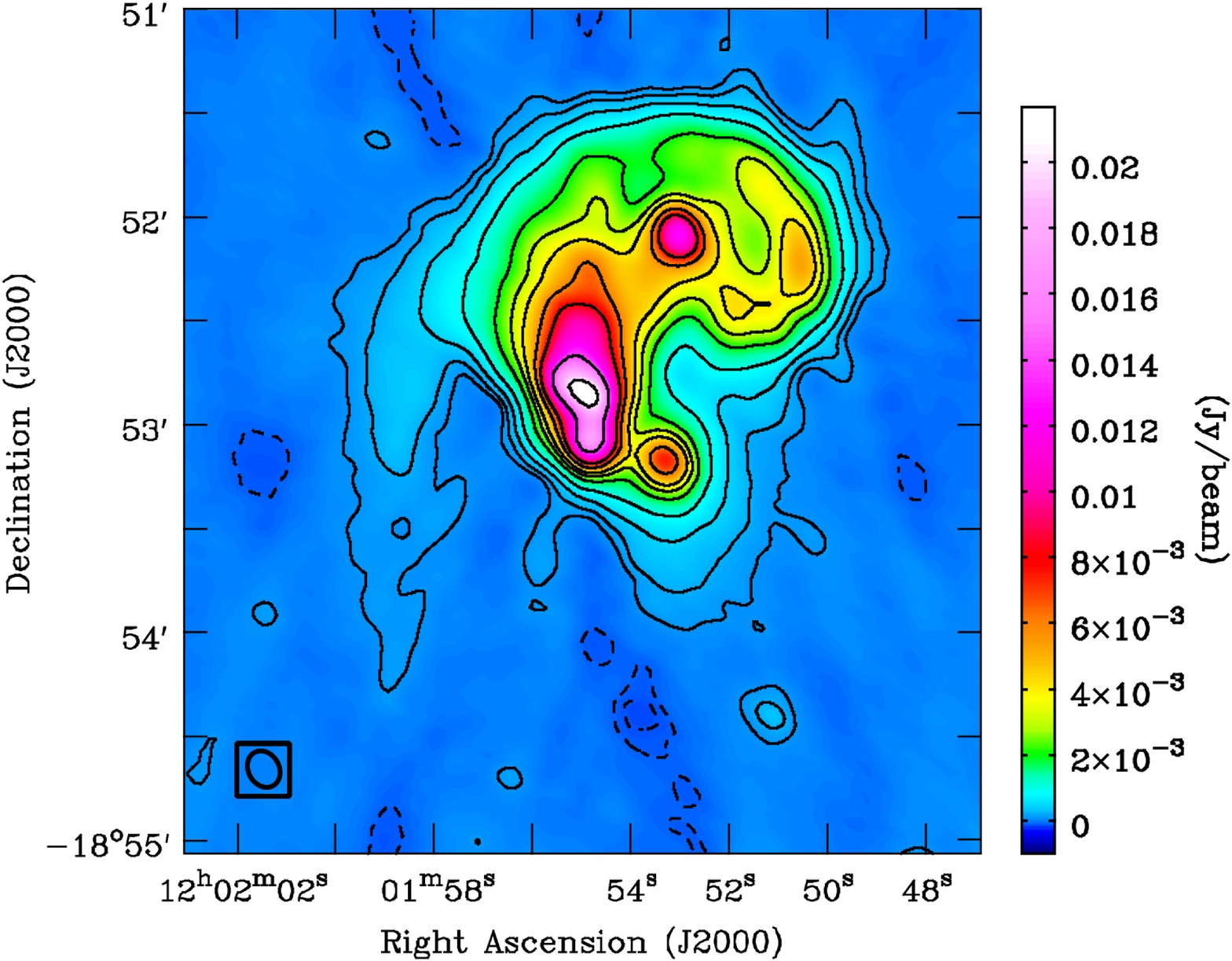}}} &
{\mbox{\includegraphics[width=8.3cm, trim=40mm 10mm 105mm 5mm,clip]{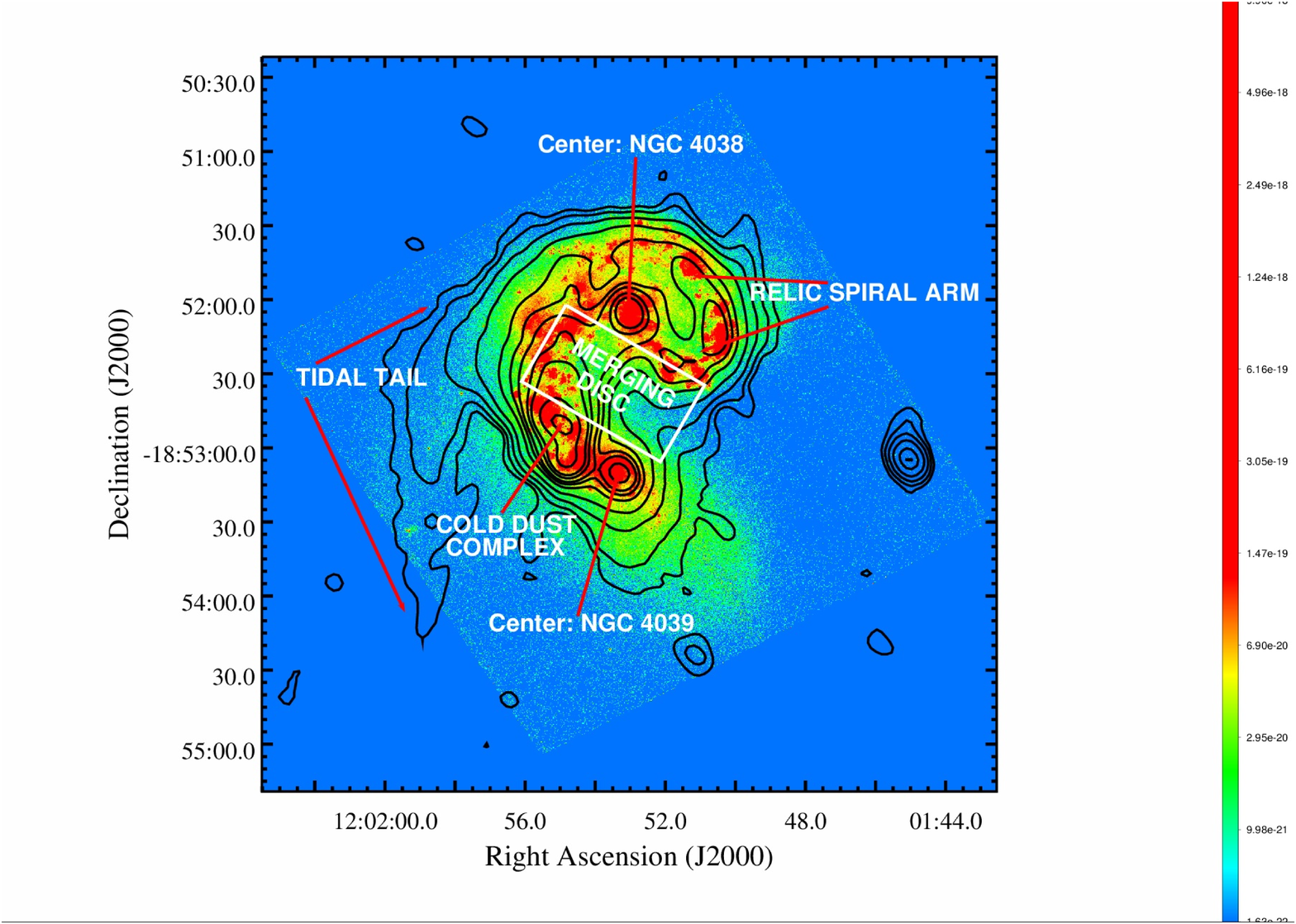}}} \\
\end{tabular}
\caption{{\it Left-hand panel:} Total intensity image of the Antennae made by
combined DnC+CnB array data at a centre frequency of 2.8 GHz having angular
resolution $11\times9$ arcsec$^2$.  Overlaid are the contours at [$-3, -2, 3,
6, 12, 24, 50, 75, 100, 150, 200, 350, 500$] $\times 40~\mu$Jy beam$^{-1}$
levels. The dashed contours represents negative total intensities. {\it
Right-hand panel:} We label the various regions in NGC4038/9 that are studied
in the paper on the HST-F658N narrow-band filter image tracing the H$\alpha$
emission of NGC 4038/9 with total radio continuum intensity contours shown same
as the left-hand panel.}
\label{totI}
\end{figure*}

In the current framework of hierarchical structure formation, galaxies build up
their mass by merging.  Galaxy encounters can compress, stretch and reshape
fields in the progenitor galaxies, hence they provide a conducive environment
for magnetic field amplification \citep{kotar10}. Since merger events were more
frequent in the early Universe \citep[e.g.,][]{patto02}, it is important to
assess how galaxy interactions affect the strength and geometry of
galactic-scale magnetic fields in order to understand the overall evolution of
cosmic magnetism.  Gravitationally interacting galaxies possess a range of
magnetic field properties.  For example, despite their irregular appearances,
the tidally interacting Magellanic Clouds have been shown to host large-scale
ordered fields of microgauss strength \citep{gaens05, mao08, mao12}.  On the
other hand, \citet{drzaz11} suggested, based on study of 16 merger pairs, that
interacting galaxies have lower field regularities and stronger total magnetic
field strengths than non-interacting ones.  Unfortunately, due to the lack of
Faraday rotation measure (RM) information, the magnetic field coherency in
these systems could not be probed.  Moreover, the large distances to their
sample galaxies and the limited angular resolution prevent one from studying
magnetic field structures on scales $< 7$ kpc.  Therefore, a high resolution
mapping of the magnetic field of a prototypical merger event is much needed.
To date, detailed high-resolution studies of galactic magnetic fields are of
individual galaxies in isolation---only a few focus on magnetism in interacting
galaxies \citep[e.g.,][]{brind92, humme95, chyzy04, rampa08}.

The Antennae pair (NGC 4038/9) is the nearest merger \citep[22 Mpc;][]{schwe08}
between two gas-rich spirals. The bodies of the colliding galaxies host sites
of active star formation in the form of super starclusters, with a global star
formation rate of 20 M$_\odot$\,yr$^{-1}$ \citep{zhang01}.  The Antennae
galaxies hosts tidal tails that measure over 100 kpc in H{\sc i} and likely
originate from outskirts of the progenitors. The Antennae have been studied
extensively from the radio to X-ray. They are also the subject of several
numerical simulations \citep[e.g.,][]{mihos93, karl10, kotar10}.  The latest
work by \citet{karl10} proposed that the progenitors first encountered
$\sim600$ Myr ago and they have just undergone the second passage.  Our
understanding of the pairs' interaction history, its multi-wavelength emission
and its proximity make the Antennae an ideal candidate for a detailed magnetic
field study.  

Magnetism in the Antennae was studied by \citet{chyzy04} with the Very Large
Array at 1.49, 4.86 and 8.44 GHz. Enhanced polarized emission is found near the
root of a tidal tail which is suggestive of a remnant spiral field.
\citet{chyzy04} computed the RM of diffuse polarized emission at 8.44 GHz and
4.86 GHz at a resolution $17\times14$ arcsec$^2$ ($\sim2$ kpc linear scale).
They pointed out that in the region where the galactic discs overlap, RMs are
coherent on the scale of several synthesized beams, likely tracing the
large-scale magnetic fields in the progenitors.  There are several regions with
consistent RM sign, which is suggestive of coherent magnetic fields. However,
these RMs could suffer from the $n\pi$ ambiguity because they were computed
using polarization angle measurements at only two bands, separated widely in
frequency. A wideband study of the diffuse polarized emission from the Antennae
is much needed to consistently derive RM to confirm the existence of coherent
magnetic fields.

In this paper, we present study of the magnetic field properties in the
Antennae galaxies. In \textsection 2, we describe our observations and data
analysis procedure. The results on Faraday depolarization and magnetic field
strengths are presented in \textsection 3 followed by discussion in
\textsection 4. Our results are summarized in \textsection 5.

\section{Observations and analysis}

We carried out wideband polarimetric observations of the Antennae galaxies
using the Karl G. Jansky Very Large Array (VLA) in the hybrid DnC array
configuration on 10-May-2013 and CnB array configuration on 09-September-2013
(project code: 13A-400). Data between 2 and 4 GHz (divided into 1024 2-MHz
channels) were recorded with the {\sc widar} correlator.  The 1024 channels
were divided in 16 sub-bands known as spectral windows.  Two scans of 15
minutes each using the DnC array and 30 minutes each using the CnB array on the
Antennae were interspersed with $\sim3$ minute scans on the phase calibrator
J1130$-$1449.  3C 286 was observed as the flux, bandpass and absolute
polarization angle calibrator for 10 minutes at the beginning of each
observation run. Unpolarized point sources, J0713$+$4349 and J1407$+$2827, were
used to calibrate polarization leakages.  We used the \citet{perle13a} flux
density scale to determine the absolute flux density of the flux calibrator 3C
286 and its absolute polarization angle was set to $+33^\circ$ across the
entire observing band \citep{perle13b}.

Data reduction was carried out using the Common Astronomy Software
Applications\footnote{http://casa.nrao.edu/} ({\sc casa}) package following
standard data calibration procedure for each array configuration separately.
The task `{\sc rflag}' was used to automatically flag data affected by radio
frequency interference (RFI).  Further manual inspection of the data was done
to remove low-lying RFI features. Overall, for both array configurations,
approximately 750 MHz of data were unusable and the remaining $\sim1200$ MHz of
data (non-continuous, spanning between 2 and 3.6 GHz) was used for further
analysis.  Several rounds of calibration and additional flagging were done
iteratively, and the gain solutions were transferred to the target source.  To
estimate the on-axis polarization leakage post calibration, we made a linearly
polarized intensity ($PI$) image of the unpolarized calibrator J0713$+$4349.
The image was consistent with noise throughout. We estimate the on-axis
instrumental polarization leakage as the ratio of the maximum value of the $PI$
image at the position of J0713$+$4349 to its total flux density, which was
found to be $<0.2$ per cent.

\begin{figure}
\begin{centering}
\includegraphics[width=8cm]{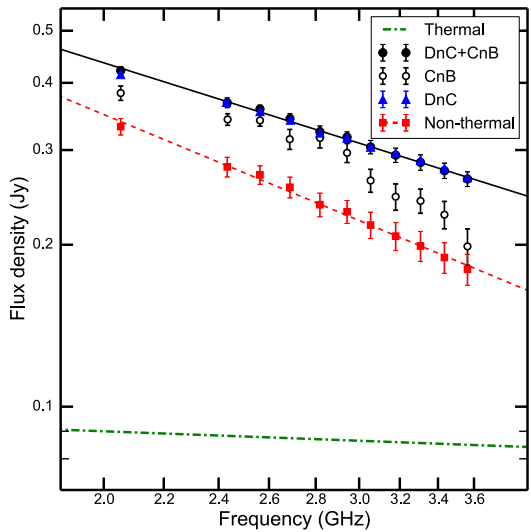}
\end{centering}
\caption{Integrated flux densities between 2 and 3.6 GHz for each spectral
window of the Antennae galaxies.  The solid black circles, open circles, and
blue triangles show the total intensity using DnC+CnB, CnB, and DnC array
configurations, respectively.  The black line is the power-law fit of a
spectral index $-0.85\pm0.02$ to the DnC+CnB array data points.  The red
squares are the non-thermal emission after subtracting the thermal emission
(green dashed-dot line). The red dashed line is the fit to the non-thermal
emission with a non-thermal spectral index $-1.11\pm0.03$.}
\label{integ}
\end{figure}

\subsection{Total intensity imaging}

We binned the 2-MHz data into 8-MHz channels for further analysis resulting in
149 8-MHz channels.  Several iterations of {\it phase only} self-calibration
were done using point sources chosen by deconvolving visibilities $\gtrsim1$
k$\lambda$ and using a uniform weighting scheme (Briggs' robust parameter =
$-2$). Self-calibration was done for each array configuration and spectral
window independently.  After satisfactory phase solutions were obtained, one
round of {\it amplitude and phase} self-calibration was carried out for a
solution interval of 15 minutes considering the entire $uv-$range.  

Finally, the calibrated data were imaged and deconvolved using the Multi-Scale
Multi-Frequency Synthesis algorithm ({\sc ms-mfs}; \citealt{rau11}) available
in the {\sc casa} package.  At this stage, we combined the {\it uv} data from
DnC and CnB arrays.  The combined image has the optimum resolution to study the
small-scale structures while being sensitive to the large-scale diffuse
emission.  To model the small, as well as the large angular scale structures,
we used six deconvolving scales varying linearly from one synthesized beam size
to $\sim1.5$ arcmin, i.e., half the angular extent of the Antennae.  The
frequency dependence was modelled using two Taylor terms ($nterms=2$).  The
final total intensity image was made with an effective bandwidth of 1.6 GHz
centered at 2.8 GHz.  Figure~\ref{totI} (left-hand panel) shows the {\it
natural weighted} total intensity DnC and CnB array combined image with an
angular resolution of $11\times9$ arcsec$^2$ \footnote{The angular resolution
corresponds to $\sim1$ kpc linear scale at the distance of the galaxies.} and
$1\sigma$ noise level of $\sim40~\mu$Jy beam$^{-1}$. The integrated flux
density of the Antennae is 336$\pm$9 mJy at 2.8 GHz. This is in good agreement
with interpolated flux densities between 1.45 and 4.85 GHz \citep{chyzy04}.
The notable features detected in the galaxies, such as the tidal tail towards
south, the central cores, the remnant spiral arm in the north and the cold dust
complex are marked in Figure~\ref{totI} (right-hand panel).

\begin{figure*}
\begin{tabular}{c}
{\mbox{\includegraphics[width=12cm, trim=0mm 10mm 130mm 10mm,clip]{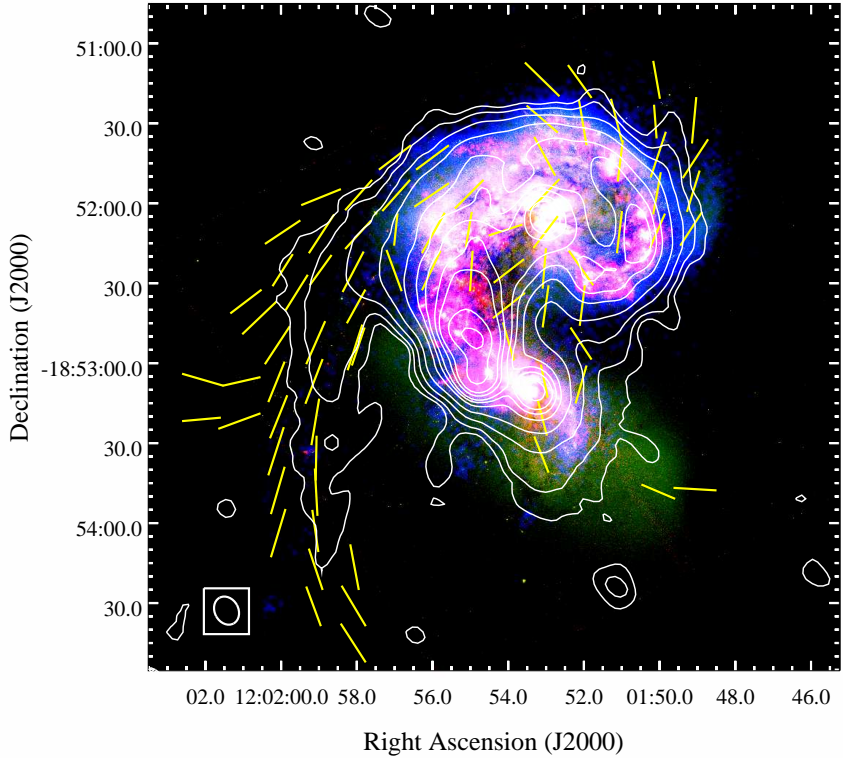}}} \\
\end{tabular}
\caption{Composite image of the Antennae with \textit{GALEX} FUV in blue,
HST-F550M \textit{y-}band image in green and HST-F658N narrow band H$\alpha$
image in red. The overlaid contours are the same as Figure~\ref{totI} and the
segments show the Faraday rotation-corrected magnetic field orientations. A 
colour version of the Figure is available in the online version.}
\label{composite}
\end{figure*}

We also made total intensity maps for each spectral window. In
Figure~\ref{integ}, we present the galaxy-integrated total flux density of the
Antennae at each of the 11 usable spectral windows for the different array
configurations: DnC, CnB and DnC+CnB.  The flux densities measured using the
CnB array (open circles) are significantly underestimated due to  missing flux
density from lack of short baselines.  However, the flux densities measured
using the DnC+CnB array (solid black circles) agree well with the flux
densities measured using the DnC array (solid blue triangles). The higher
resolution DnC+CnB array images are sensitive to both small-scale as well as
large-scale diffuse emission and hence we use this image for the rest of our
analysis. 

Radio continuum emission in galaxies mainly originates from non-thermal
synchrotron and thermal free--free emission. In order to study the magnetic
field properties in galaxies, contribution from the thermal emission needs to
be separated from the total radio emission. We have used star formation rate,
estimated via extinction corrected far ultraviolet (FUV) emission, as the
tracer of thermal emission. A detailed description of the thermal emission
separation method is given in Appendix~\ref{thermal}. We note, that the
estimated thermal emission can suffer from systematic errors up to $\sim30$ per
cent in regions of high dust extinction or starbursts. However, the errors in
the estimated non-thermal emission in those regions are less than $\sim20$ per
cent. The estimated thermal emission is shown as the green dashed-dot line in
Figure~\ref{integ}.  The red squares shows the non-thermal emission after
separating the thermal emission at each spectral window and is well fitted by a
power-law (red dashed line) with spectral index $-1.11\pm0.03$. However, due to
uncertainties in the estimated thermal emission, there can be systematic error
up to $\sim10$ per cent on the value of the spectral index.

\subsection{Rotation measure synthesis}

The plane of polarization of a linearly polarized signal is rotated when it
passes through a magneto-ionic medium because of the Faraday rotation
effect. The angle of rotation depends on the Faraday depth ($\phi$) and
is given by,
\begin{equation}
\left(\frac{\phi}{\rm rad~m^{-2}}\right) = 0.812 \int_{\rm source}^{\rm observer} \left(\frac{n_e(l)}{\rm cm^{-3}}\right) \left(\frac{B_{\rm \parallel}(l)}{\rm \mu G}\right)~\left(\frac{dl}{\rm pc}\right).
\end{equation}
Here, $n_e$ is the density of thermal electrons, $B_{\parallel}$ is the
magnetic field component along the line of sight and $dl$ is the path length
through the magneto-ionic media. We employ the technique of RM synthesis
\citep{brent05} to estimate $\phi$.

To facilitate a spatially resolved Faraday rotation study across NGC 4038/9,
Stokes $Q$ and $U$ images of the combined DnC+CnB array for each of the 149
8-MHz channels were made.  We applied the natural weighting scheme to the $uv$
data and performed multi-scale {\sc clean}. Since the angular resolution for
each channel is different, we convolved all the channel maps to a common
resolution of $15\times12$ arcsec$^2$ (resolution of the lowest frequency
channel) before combining them into a single image cube.  RM synthesis and
deconvolution of the $\phi$ spectrum were performed on this image cube using
the {\sc pyrmsynth} package \citep{bell13}.  The typical rms noise in the
Faraday depth spectrum is $\sim10~\mu$Jy beam$^{-1}$.  We cleaned down
to 3$\sigma$ ($\sim30~\mu$Jy beam$^{-1}$) when deconvolving the pixel-by-pixel 
Faraday depth spectrum.

The $\lambda^2$ coverage determines the maximum observable $\phi$ and the
sensitivity to extended $\phi$ structures. For our data, the maximum observable
$\phi$ ($|\phi_{\rm max}|$) is $\sim1\times 10^4~\rm rad~m^{-2}$ and our
sensitivity to extended structures in $\phi$ drops to 50 per cent at $450~\rm
rad~m^{-2}$.  The RM spread function (RMSF) was computed from the $\lambda^2$
coverage by weighting each frequency channel by their noise. The RMSF for our
observations has a full-width at half maximum (FWHM) of $219 \rm ~rad~m^{-2}$.

Initially, we performed a low-resolution search for significant $\phi$
components in the range $-2\times10^4 \rm ~rad~m^{-2}< \phi <2\times10^4
~rad~m^{-2}$ in steps of 50 rad m$^{-2}$. No significant component was found
for $|\phi|>10^3~\rm rad~m^{-2}$.  Then, we performed a higher resolution
search in the range $-2000 \rm ~rad~m^{-2}< \phi <2000~rad~m^{-2}$ in steps of
10 rad m$^{-2}$ to oversample the FWHM.  We considered five adjacent values
around the peak and fitted the peak to a parabola to determine the peak Faraday
depth and the corresponding peak polarized intensity. To reduce the effect of
the Ricean bias due to positive noise background of the polarized intensity
map, we considered only those pixels where the peak polarized intensity was
more than 7$\sigma$ level.  We therefore do not correct for the Ricean bias as
its effect would be less than 3 per cent for polarized intensity $>7\sigma$
\citep{wardl74}.  

Foreground contribution to the Faraday depth from the Milky Way was estimated
using the \citet{opper12} Galactic RM map. The foreground RM was found to be
$-28\pm7$ rad m$^{-2}$ in the direction of the Antennae.  We adopt $-30~\rm
rad~m^{-2}$ as the contribution from the Milky way \citep[similar
to][]{chyzy04} to the $\phi$ obtained through RM synthesis.

\section{Results}

\subsection{Total radio intensity}

In the left-hand panel of Figure~\ref{totI} we present the total intensity map
of NGC 4038/9 across the frequency range 2 to 3.6 GHz.  In the right-hand panel
we label the prominent features visible in our total intensity images.  The
notable features are as follows: (1) the central cores of the two galaxies, (2)
the remnant spiral arm of the northern galaxy NGC 4038, (3) the tidal tail
toward the south and (4) the cold dust complex located northeast of the
southern galaxy NGC 4039. These features were also visible in the 4.86 GHz
observations by \citet{chyzy04}.

In Figure~\ref{composite}, we show a composite image of the Antennae with
{\it GALEX} far-ultraviolet (FUV) in blue, $y-$band optical image in green
observed with the HST-F550M medium band filter, and narrow band H$\alpha$ image
in red observed with the HST-F658N narrow band filter.  The total intensity
contours at 2.8 GHz and the line segments of magnetic field vectors corrected
for Faraday rotation are overlaid.  The local peaks of the radio emission
closely follow the sites of star formation traced by the H$\alpha$ and FUV
emission.  The non-thermal spectral index\footnote{The non-thermal spectral
index, $\ant$, is defined as $S_\nu\propto \nu^{\alpha_{\rm nt}}$. The method
of estimating $\ant$ is discussed in Appendix \ref{spind_distr}} ($\ant$) in
the star-forming regions is comparatively flatter than in the diffuse regions
and typically lies in the range $-0.6$ to $-0.8$ close to the injection
spectral index of CREs. This indicates the radio emission originates from
freshly generated cosmic ray electrons (CREs).  The peak in the radio emission
is co-incident with the dark cloud complex in the southern part (see
Figure~\ref{totI}).  The non-thermal spectrum is flattest in this region with
$\ant\sim-0.6$ indicating that this is an efficient site for producing cosmic
ray particles.  The thermal free--free emission in this region is the brightest
although, the H$\alpha$ and FUV emission in this region are relatively weak,
likely due to high dust extinction.   

\begin{figure}
\begin{centering}
\begin{tabular}{c}
{\mbox{\includegraphics[width=9cm, trim=0mm 6mm 0mm 0mm, clip]{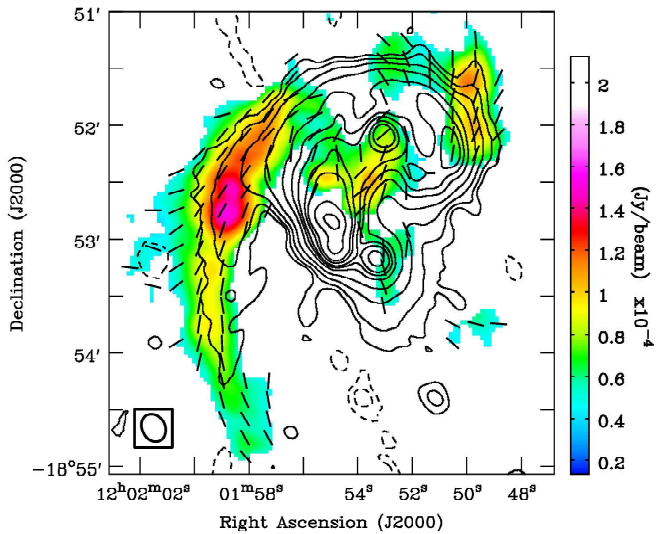}}}\\
\end{tabular}
\end{centering} 
\caption{Polarized intensity at the peak of the Faraday depth spectrum across
the Antennae. Only the pixels with peak polarized intensity $>7\sigma$ in
Faraday depth spectrum are shown and are not corrected for the Ricean bias.
Overlaid line segments indicate the Faraday rotation-corrected magnetic field
orientations.  The total intensity contours are the same as in
Figure~\ref{totI}.}
\label{polimap}
\end{figure}

\subsubsection{Southern tidal tail}

A notable feature in the total intensity map is the detection of the gas-rich
tidal tail toward the south. The northern tail has poor gas content
\citep{hibba01} and remains undetected in our observations. The southern tail
extends up to $\sim2.8$ arcmin in size, corresponding to a projected length of
$\sim18$ kpc from the base of the tail\footnote{We define the base of the tail
to be located eastward of the merging disc region at RA = $\rm
12^{h}01^{m}57.5^{s}$ and Dec. = $-18^\circ52^\prime06^{\prime\prime}$
(J2000).} at the 3$\sigma$ level.    The H{\sc i} emission in this tail spans
$\sim65$ kpc \citep{hibba01} and the detectable radio continuum emission
closely follows it up to $\sim18$ kpc.

Although the southern tail has H{\sc i} column densities that exceeds $10^{20}$
cm$^{-2}$, there is little evidence of active star formation as revealed by the
UV emission \citep{hibba05}. The tail is, however, visible in the wide band
optical images and mid-infrared images between $2-8~\mu$m,  and the UV emission
predominantly arises from stars older than the dynamical age of the tidal tail
\citep{hibba05}.  The FUV emission from the southern tail is weak and
corresponds to an average star formation rate of $\sim10^{-3}~\rm
M_\odot~yr^{-1}$ over the entire radio extent. This FUV emission could also
arise from less massive ($\lesssim5~\rm M_\odot$) and older stellar populations
that were stripped from the galactic disc due to the interaction. This suggests
that the CREs giving rise to the radio emission were produced in the
star-forming disc before the first interaction $\sim6\times10^8$ years ago
\citep{karl10}. In \textsection3.4, we estimate the magnetic field
strength in the tidal tail to be $\sim10~\mu$G assuming energy equipartition
between cosmic ray particles and magnetic field \citep{beckkrause05}. The
synchrotron lifetime of the CREs emitting at 2.8 GHz in a magnetic field of
$10~\mu$G is $\sim3\times10^7$ years, which is shorter than the dynamical age
of the tail. Therefore, the CREs in the tidal tail are composed of relatively old
population which gives rise to steep non-thermal spectrum with $\ant$ in the
range $-1.2$ to $-1.6$ (see Figure~\ref{spind_distr}).

\begin{figure*}
\begin{centering}
\begin{tabular}{cc}
{\mbox{\includegraphics[width=9.2cm, trim=0mm 0mm 0mm 0mm, clip]{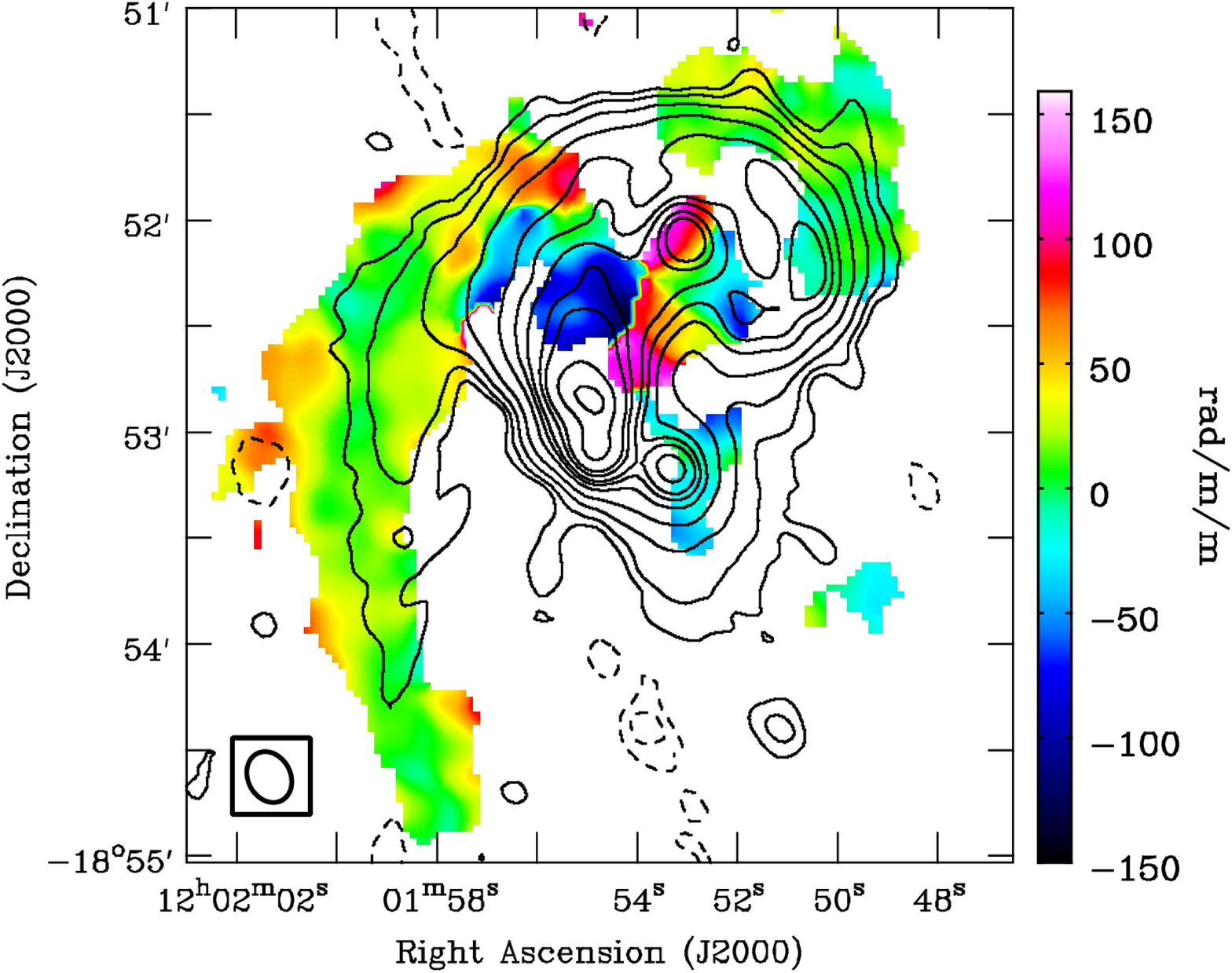}}} &
{\mbox{\includegraphics[width=9cm, trim= 0mm 0mm 5mm 5mm, clip]{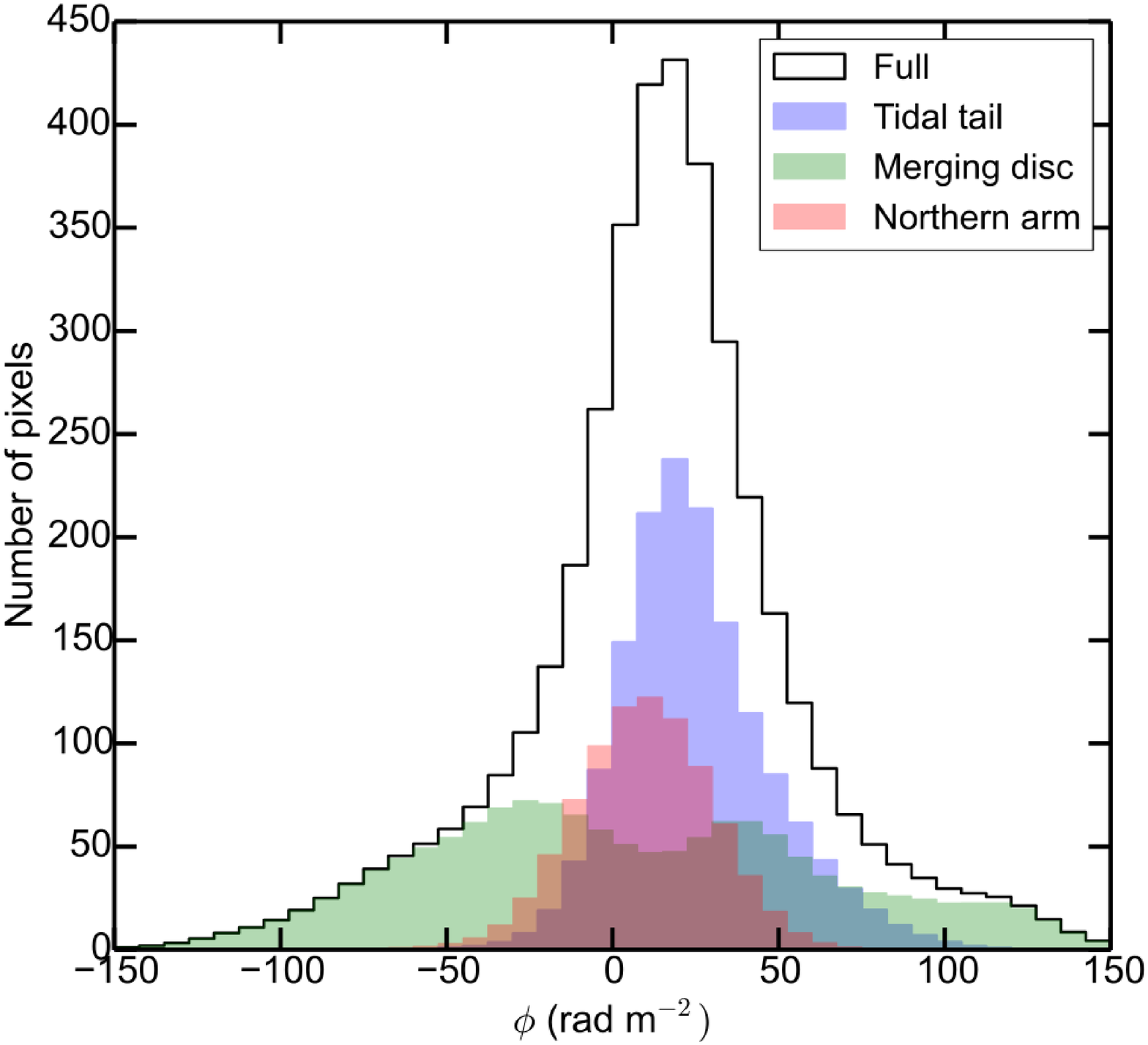}}}\\
\end{tabular}
\end{centering}
\caption{{\it Left-hand panel}: Faraday depth map of the Antennae computed
using RM synthesis overlaid with the total intensity contours same as in
Figure~\ref{totI}.  A foreground Milky Way contribution of $-30\rm ~rad~m^{-2}$
has already been subtracted.  {\it Right-hand panel}: The distribution of
Faraday depth in various regions in the Antennae labelled in Figure~\ref{totI}.
A colour version of the Figure is available in the online version.}
\label{rmmap}
\end{figure*}

\begin{figure*}
\begin{centering}
\begin{tabular}{ccc}
{\mbox{\includegraphics[height=6cm, trim=10mm 10mm 10mm 10mm, clip]{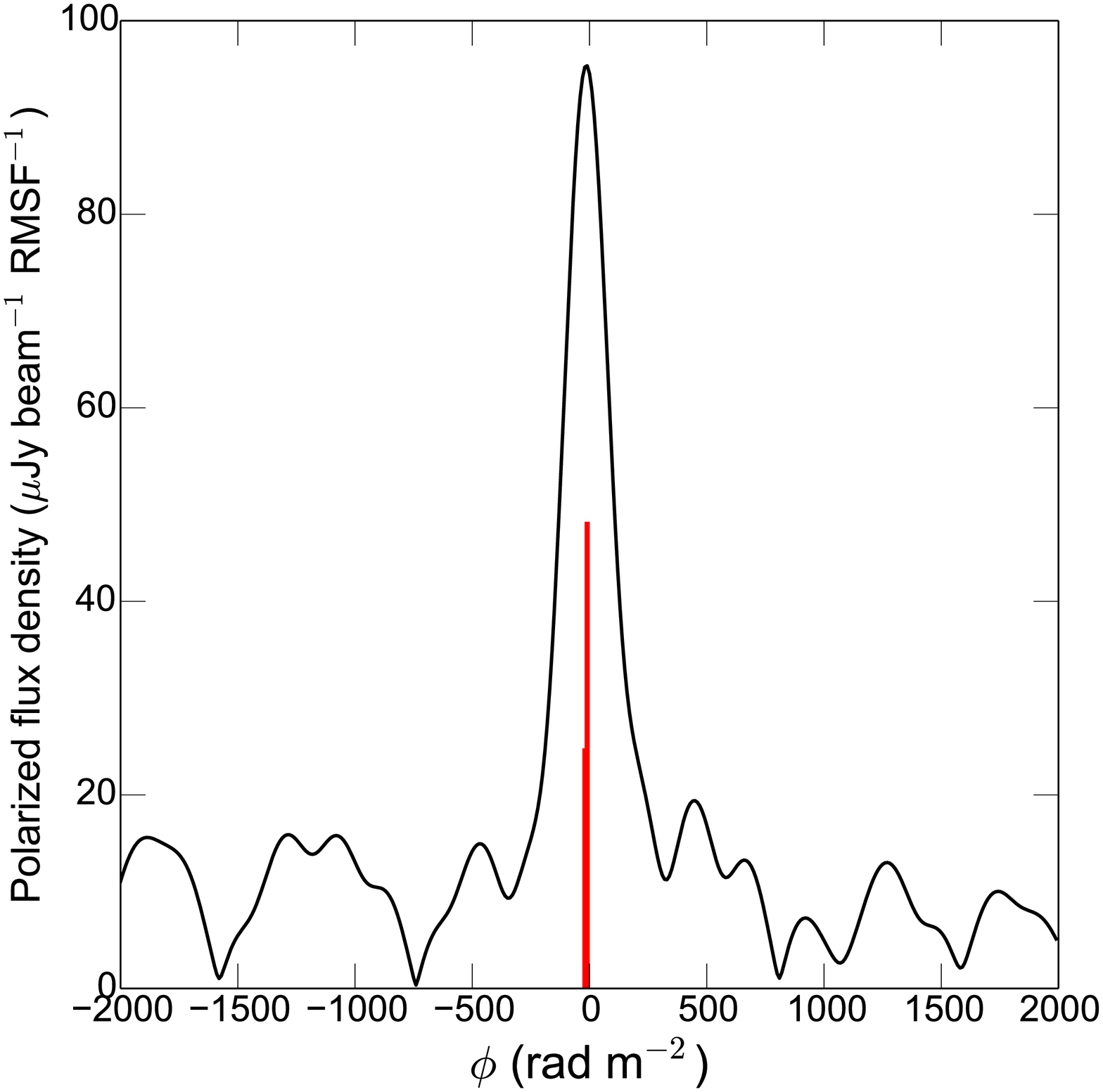}}}&
{\mbox{\includegraphics[height=6cm, trim=10mm 10mm 10mm 10mm, clip]{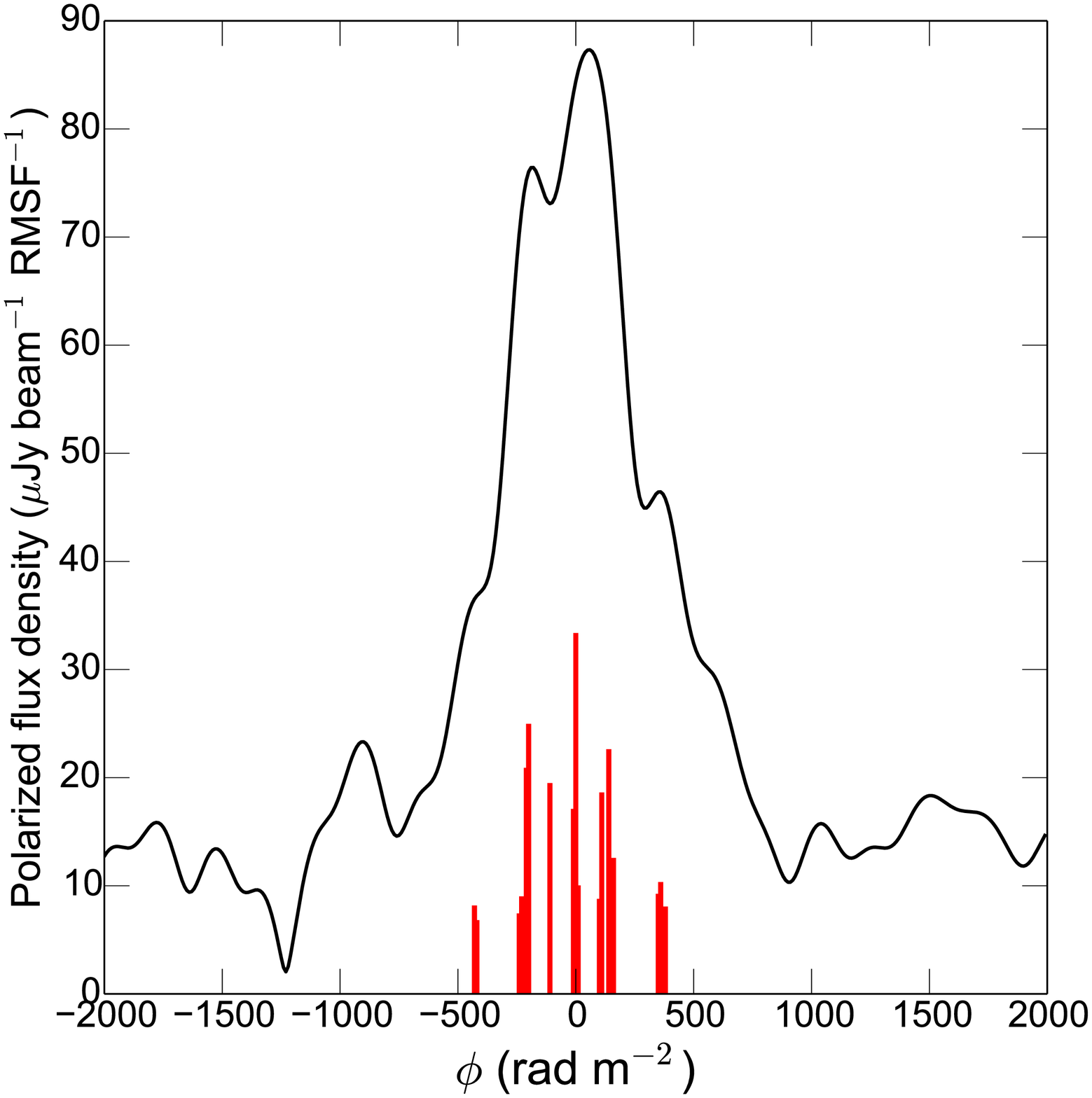}}}&
{\mbox{\includegraphics[height=6cm, trim=10mm 10mm 10mm 10mm, clip]{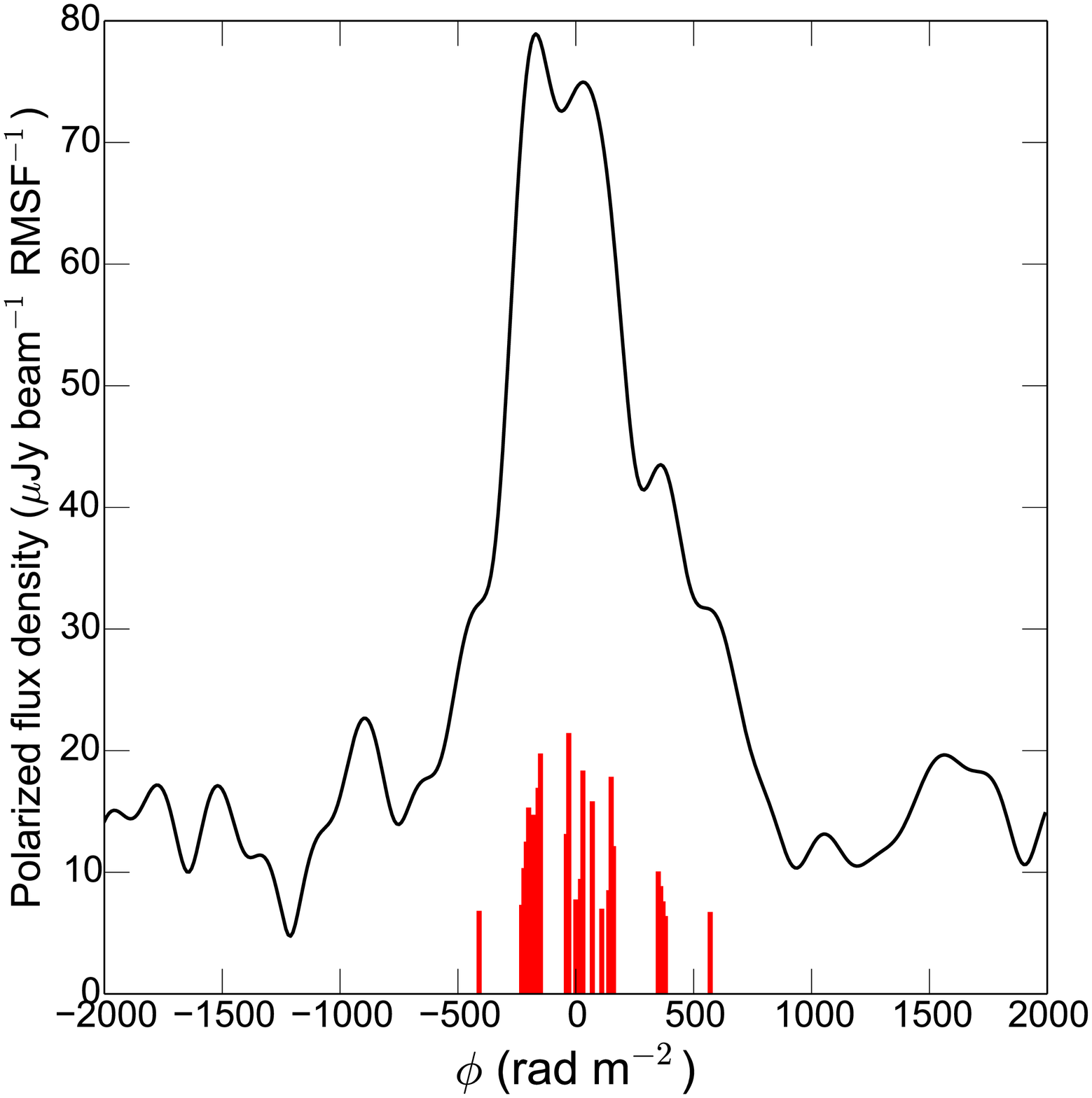}}}\\
\end{tabular}
\end{centering}
\caption{Cleaned Faraday depth spectra (black curves) at different locations in
the Antennae uncorrected for the Milky Way contribution. The red lines are the
Faraday depth clean components. {\it Left-hand panel:} The typical Faraday
depth spectra along the tidal tail with a simple profile. The Faraday depth
components lies within the step size (10 rad m$^{-2}$) of performing the RM
synthesis.  {\it Middle and right-hand panel:} The Faraday depth spectra of two
adjacent pixels in the merging disc region which shows complicated Faraday
depth spectra with several Faraday depth components.} 
\label{spectra}
\end{figure*}

\subsection{Polarized intensity and Faraday depth}

The polarized intensity image of the Antennae galaxies at 2.8 GHz
determined from the peak of the Faraday depth spectrum is shown in
Figure~\ref{polimap}. Overlaid segments are the Faraday rotation-corrected
magnetic field orientations.  Due to our stringent cut of 7$\sigma$ on the peak
polarized intensity, we do not detect polarized emission from the entire
radio-emitting region. However, we do detect the southern tidal tail, the
merging disc and the northern part of NGC 4038.  Strongly polarized emission is
observed from the tidal tail with median fractional polarization\footnote{The
fractional polarization is defined as, $\Pi=PI/I_{\rm nt}$, where, $PI$ is the
linearly polarized intensity and $I_{\rm nt}$ is the synchrotron intensity.  We
express $\Pi$ both in terms of percentage and fraction.}, $\Pi\sim 25$ per cent
at 2.8 GHz.  The polarized intensity spans a projected linear size of $\sim20$
kpc, longer than the extent of the detectable total intensity
emission.\footnote{$\Pi$ is computed within overlapping regions in $I_{\rm
nt}$ and $PI$.  We considered only the pixels with $I_{\rm nt}\gtrsim 3\sigma$
and $PI\gtrsim7\sigma$.} The magnetic field vectors are well aligned along the
entire length of the tail. In the region of the merging discs, the polarized
emission is weakly polarized (median $\Pi\sim1.6$ per cent) and the magnetic
field is more randomly oriented.  The polarized emission and the magnetic field
vectors in the northern galaxy NGC 4038 follow a spiral pattern, however, this
emission is offset outward (to the west) with respect to the remnant spiral
arm. The peak of the polarized emission is shifted outward by $\sim1.6$ kpc
from the peak of the total intensity.  Such an offset of the polarized emission
was also seen in the 4.85 GHz observations of \citet{chyzy04}.  The polarized
emission along the peak of the total radio continuum emission in the remnant
spiral arm remains undetected. This region also hosts sites of intense star
formation and hence the polarized emission is likely depolarized due to
turbulent magnetic fields.

We did not detect any polarized emission at 2.8 GHz from the southern part of
the galaxies in general, in particular around the dark cloud complex. At
7$\sigma$ cutoff, this corresponds to a $\Pi\lesssim0.2$ per cent, comparable
to the instrumental leakage. This region shows strong Faraday depolarization
between the 4.86 and 8.44 GHz observations of \citet{chyzy04}. Moreover, they
found the percentage polarization to be $\lesssim1$ per cent at 8.44 GHz.
Since, this region also hosts strong thermal emission, the high $n_e$ would
lead to high Faraday depth. Therefore, the low fractional polarization can be
caused by either strong wavelength$-$($\lambda-$)dependent depolarization or by
$\lambda-$independent depolarization due to turbulent magnetic fields within
the beam.

\begin{figure*}
\begin{centering}
\begin{tabular}{cc}
{\mbox{\includegraphics[width=9cm]{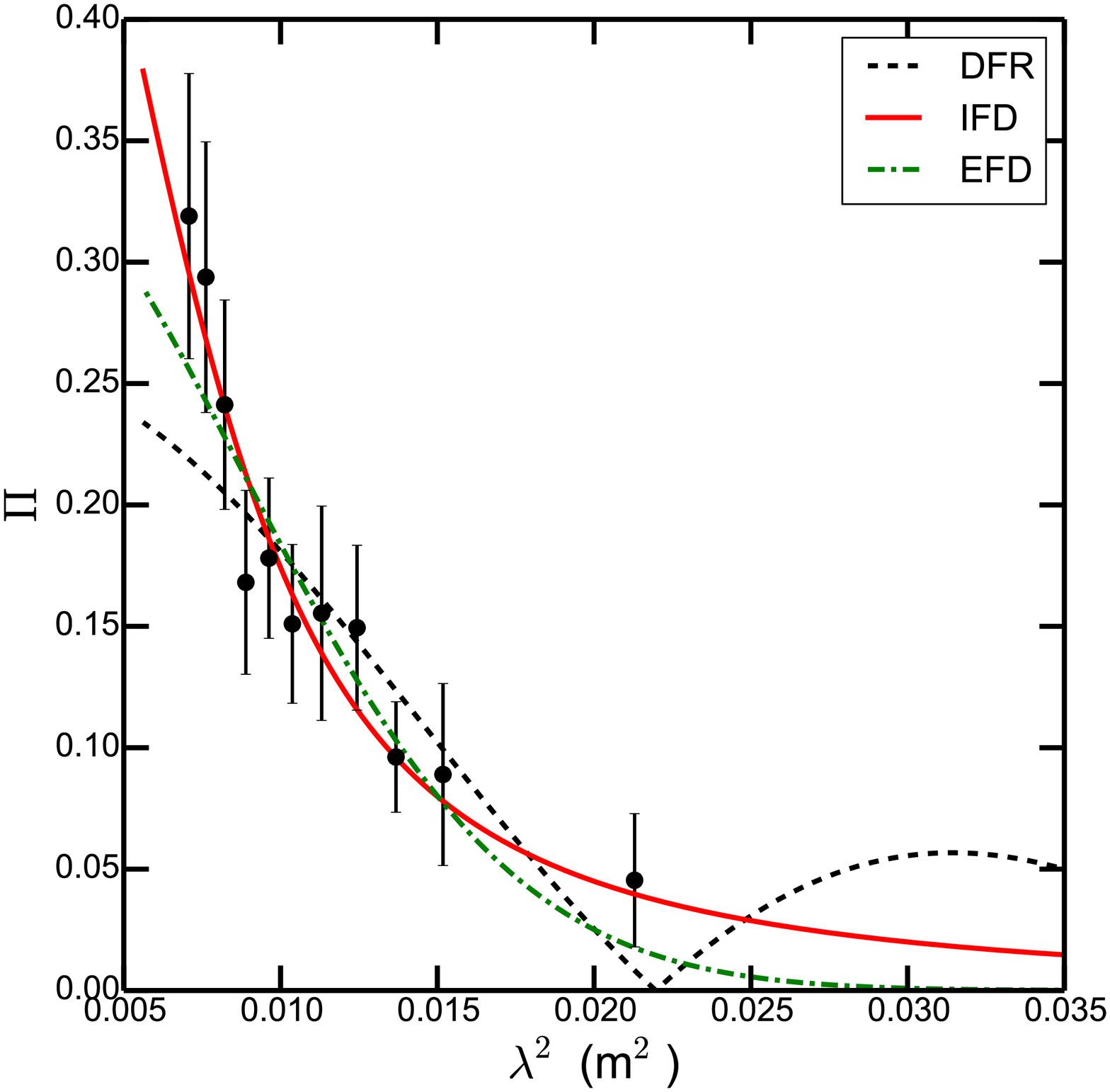}}}&
{\mbox{\includegraphics[width=9cm]{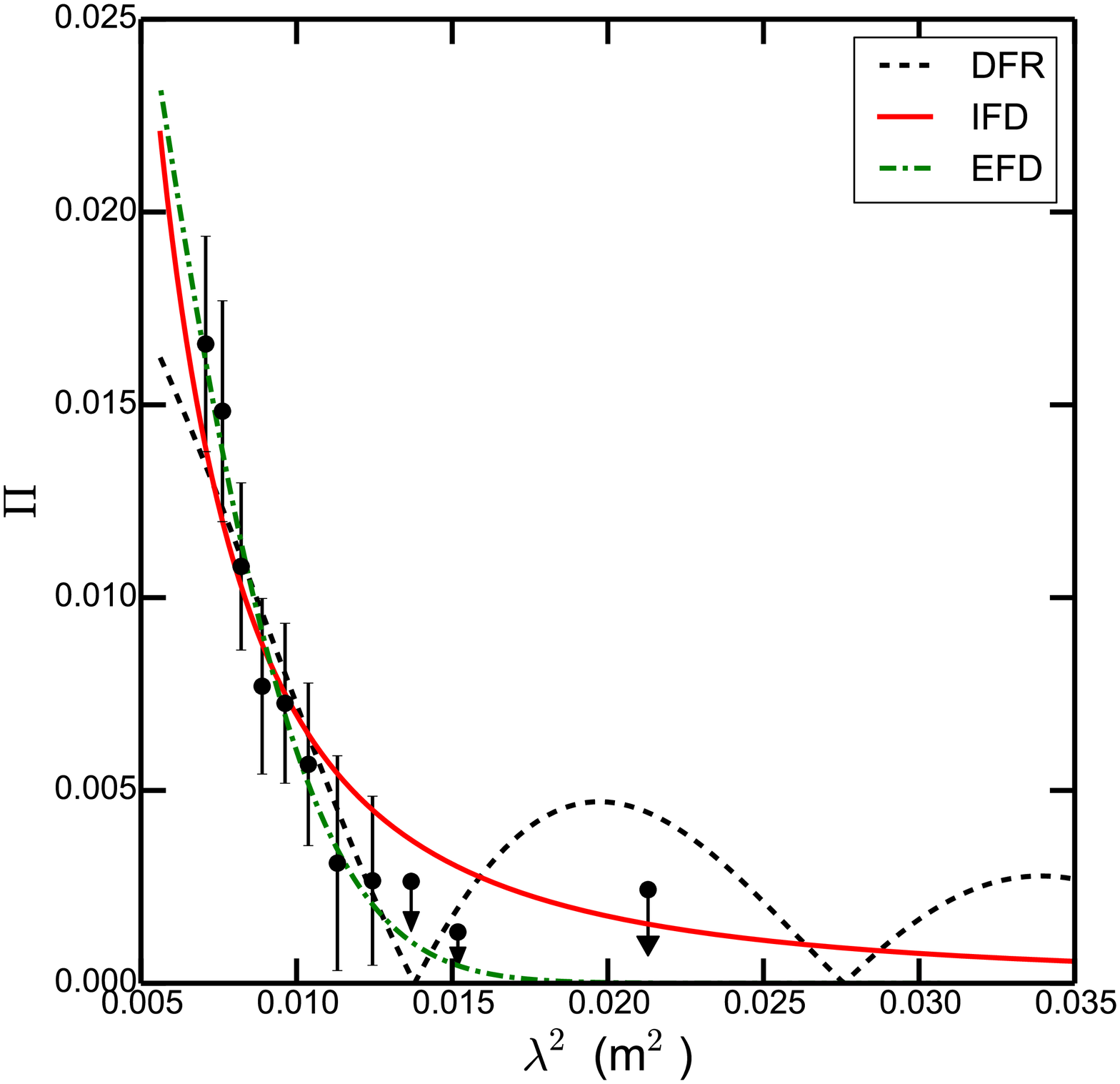}}}\\
\end{tabular}
\end{centering}
\caption{Variation of the fractional polarization ($\Pi$) with $\lambda^2$ in
the tidal tail ({\it left-hand panel}) and the merging region ({\it right-hand
panel}). The points with arrows are the 2$\sigma$ upper limits. The lines are
the fitted models of different types. For the tidal tail, the IFD model provides
the best fit and in the merging disc region, EFD model fits the data best.} 
\label{banddepol}
\end{figure*}

In left-hand panel of Figure~\ref{rmmap}, we show the Galactic
foreground-corrected map of the Faraday depth of the Antennae and its
pixel-wise distribution in the right-hand panel. Faraday depth along the tidal
tail vary smoothly having a mean of $+25$ rad m$^{-2}$ and a standard deviation
of 22 rad m$^{-2}$ (shown as the blue histogram) measured over several beams in
the plane of the sky. In the northern polarized arm, we find the mean Faraday
depth to be $+9.3$ rad m$^{-2}$ with dispersion of 20.5 rad m$^{-2}$ (shown as
the red histogram).  The dispersion of Faraday depth around its mean value is
larger than that of the tidal tail, suggesting the magnetic field is more
ordered in the tidal tail.

The merging disc has strong $\phi$ variations in the range $-100$ to $+140$ rad
m$^{-2}$. The pixel-wise distribution of $\phi$ in this region shows two peaks
(shown as the green histogram in Figure~\ref{rmmap}) at $\sim-25$ rad m$^{-2}$
and $\sim+40$ rad m$^{-2}$.  The Faraday depth in this region is observed to
change sign abruptly and is co-spatial with the systemic velocity of the H{\sc
i} emission. The Faraday depth spectra in this region have complicated
structures having highly dispersed $\phi$ components as compared to other
regions. The middle and the right-hand panels of Figure~\ref{spectra} show the
Faraday depth spectra at two adjacent pixels where the sign change is observed.
Clearly, the spectra are not single peaked as compared to the spectra in the
tail (left-hand panel).  For the middle panel, the Faraday depth spectrum has a
peak at $+68$ rad m$^{-2}$, while for the right-hand panel, the Faraday depth
spectrum peaks at $-130$ rad m$^{-2}$. The restoration of the multiple
clean $\phi$ components (shown as red lines in Figure~\ref{spectra}) with the
RMSF of 219 rad m$^{-2}$ gives rise to complicated, broad Faraday depth
spectrum. The merging region possibly hosts extended Faraday depth structures,
however it is not clear whether the multiple components are indeed different
Faraday depth components or one (or multiple) broad components. Therefore, it
is not clear if the sign change we observe in Faraday depth is real.  Both high
angular and Faraday depth resolution are needed along with larger wavelength
coverage at lower frequencies to study Faraday depolarization and the nature of
the magneto-ionic medium of the merging disc.

\subsection{Faraday depolarization}

The maximum fractional polarization ($\Pi_{\rm max}$) arising due to
synchrotron emission is given by $\Pi_{\rm max} = (1-\ant)/(5/3-\ant)$ and lies
in the range $0.7-0.75$ for $\ant$ in the range $-0.7$ to $-1.1$.  However, we
seldomly observe this theoretical maximum because the polarized emission can be
depolarized due to: (1) $\lambda-$independent beam depolarization caused by
random magnetic fields on scales smaller than the beam, (2) bandwidth
depolarization caused by Faraday rotation within a finite bandwidth and/or (3)
$\lambda-$dependent depolarization depending on the nature of the magneto-ionic
medium \citep[see e.g.,][]{sokol98, sulli12}.  Studying the nature of the
$\lambda-$dependent depolarization can help us to determine the properties of
the magneto-ionic medium.  However note, the signal-to-noise ratio (SNR) per
8-MHz channel is insufficient to estimate the polarized intensity.  Hence, we
averaged over each spectral window of 108 MHz width, after flagging the edge
channels. 

To improve the SNR further, we studied the $\lambda-$dependent depolarization
averaged over the tidal tail, the merging disc and the northern region of NGC
4038.  To account for beam depolarization while averaging the extended emission
over the region of interest, we computed $\Pi$ by averaging the pixels in
Stokes $Q$ and $U$ maps as $\Pi = \sqrt{\langle Q \rangle^2 + \langle U
\rangle^2}/\langle I_{\rm nt}\rangle$. We find evidence of strong
$\lambda-$dependent depolarization in all parts of the Antennae.  Due to
limited signal-to-noise ratio in individual channel maps, detailed fitting of
the Stokes $Q$ and $U$ {\it versus} $\lambda^2$ as done for the galaxy M 51 by
\citet{mao15} was not possible. Hence, we fitted the fractional polarization as
a function of $\lambda^2$ using the simple models given in \citet{sulli12}
based on \citet{sokol98}.

We note that the depolarization\footnote{Depolarization is defined as the ratio
of polarization fraction at $\nu_1$ and $\nu_2$, where $\nu_1 < \nu_2$.}
between 2 and 3.6 GHz is $\sim0.15$.  For this depolarization to arise due to
bandwidth depolarization because of averaging over 108 MHz, $|\phi|\gtrsim
10^3$ rad m$^{-2}$ is required. The maximum observed $\phi$ is $\sim150$ rad
m$^{-2}$ and thus bandwidth depolarization is not the dominant depolarization
mechanism.  Therefore, we do not consider this effect in our future
calculations.

\subsubsection{Tidal tail and merging disc} \label{depoltailmerger}

To assess which mechanism causes the $\lambda-$dependent depolarization, we
model the polarization fraction as a function of $\lambda^2$ using the
different depolarization models given in \citet{sokol98}: differential Faraday
rotation (DFR), internal Faraday dispersion (IFD), and external Faraday
dispersion (EFD). In Figure~\ref{banddepol}, we show the variation of the $\Pi$
with $\lambda^2$ computed within each spectral window of $\sim108$ MHz width.
The left-hand panel is for the tail region and the right-hand panel is for the
merging disc region. 

For the tidal tail region, $\Pi(\lambda)$ is best fitted by the IFD model (red
solid line in Figure~\ref{banddepol} left-hand panel) with $\chi^2=3.2$ as
compared to $\chi^2$ of 10.3 and 5.7 for the DFR and EFD models, respectively.
As per the IFD model, the Faraday-rotating medium contains turbulent magnetic
fields along the line-of-sight and is also emitting synchrotron radiation. In
this case $\Pi(\lambda)$ varies as
\begin{equation}
\Pi(\lambda) = \Pi_{\rm int} \left(\frac{1-e^{-S}}{S}\right).
\label{eqnIFD}
\end{equation}
Here, $\Pi_{\rm int}$ is the $\lambda-$independent intrinsic polarization
fraction, and $S=2\sigma_{\rm RM}^2\lambda^4$, where $\sigma_{\rm RM}$ is the
Faraday dispersion within the 3D beam arising from turbulent fields.  We find
$\sigma_{\rm RM} = 131\pm23$ rad m$^{-2}$ and internal polarization fraction of
$0.62\pm0.18$ close to the theoretical value. 

Similarly, for the merging disc region, $\Pi(\lambda)$ is best fitted by an
external dispersion screen (green dashed-dot line in Figure~\ref{banddepol}
right-hand panel) with $\chi^2=1.5$. The fits with DFR and IFD models
gives $\chi^2$ of 4.8 and 6.1, respectively. In an external dispersion screen,
the Faraday rotating medium, lying between the observer and the synchrotron
emitting media, contains random magnetic field along the line of sight. In
this case $\Pi(\lambda)$ is given by
\begin{equation}
\Pi(\lambda) = \Pi_{\rm int} e^{-2\sigma_{\rm RM}^2\lambda^4}.
\label{eqnDFR}
\end{equation}
We find the internal polarization fraction to be $0.040\pm0.006$ and
$\sigma_{\rm RM}=99\pm5$ rad m$^{-2}$.

\subsubsection{Star-forming regions}\label{sfreg}

We did not detect any polarized emission from the dark cloud and star-forming
regions located in the southern part of NGC 4038/9. While the synchrotron
emission is the strongest in these regions, non-detection of polarized emission
at the resolution of our observations (corresponding to $\sim1$ kpc linear
scales) suggests the magnetic field could be random at much smaller scales.
However, we could estimate the effects of Faraday depolarization in these regions
using the estimated thermal emission. The free--free optical depth ($\tau_{\rm
ff}$) at a radio frequency ($\nu$) is related to the thermal flux density
($S_{\rm th,\nu}$) as $S_{\rm th,\nu} = 2kT_e\nu^2 \Delta\Omega
(1-e^{-\tau_{\rm ff}})/{c^2}$. Here, $\Delta\Omega$ is the solid angle
subtended by the source and $T_e$ is the electron temperature assumed to be
$10^4$ K. The $\tau_{\rm ff}$ is related to the emission measure
($\textrm{EM}$) as
\begin{equation}  \label{tauff}
\tau_{\rm ff} = 0.082 ~T_e^{-1.35} \nu^{-2.1} \textrm{EM}.
\end{equation}
Here, $n_e$ is the thermal electron density and is related to the $\textrm{EM}$
as $n_e \approx (\textrm{EM}~f/h_{\rm ion})^{1/2}$, where $f$ is the filling
fraction. In a star-forming disc, the $\textrm{EM}$ predominantly arises from
the H{\sc ii} regions with small filling factors, $f\sim5$ per cent
\citep{ehle93, beck07}.\footnote{We note that the filling factor is a
difficult quantity to measure in external galaxies. We therefore use
representative values within a range of factor of 2 in our calculations.}
$h_{\rm ion}$ is the line-of-sight depth of the ionized gas assumed to be the
linear size of the star-forming regions and is observed to have similar size as
the resolution of our observations ($\sim 1$ kpc) as seen in the H$\alpha$
images.  Using the thermal emission, we estimate the typical $n_e$ to be
$\sim3-5$ cm$^{-3}$.  Thus, the $|\phi|$ for a regular magnetic field of
$\sim1$ kpc scale having strength $|B_{\parallel}| \sim 3-5~\mu$G (see
\textsection~\ref{bord}) is $\sim1-2 \times 10^4~\rm rad~m^{-2}$.

For $|\phi|\approx2\times10^4$ rad m$^{-2}$, the effect of bandwidth
depolarization is severe below 1.5 GHz within a channel width of 8 MHz. This
effect is less than 20 per cent at 3 GHz and our observations are sensitive to
$|\phi|\sim10^4$ rad m$^{-2}$. Thus, our estimated $|\phi|$ of $\sim1-2\times
10^4$ rad m$^{-2}$ in the southern part of NGC 4038/9 may not depolarize the
emission at the higher frequency end of our observations. Moreover, the
fractional polarization in these regions was found to be $\sim1$ per cent at
8.44 GHz \citep{chyzy04}, where the effect of bandwidth depolarization is even
lower.  Thus, owing to the extreme environment and turbulence driven by super
starclusters, we conclude that the effects of Faraday depolarization could be
severe.

\subsection{Magnetic field properties}

\subsubsection{Total magnetic field strength}

\begin{figure}
\begin{tabular}{c}
{\mbox{\includegraphics[width=9cm, trim=0mm 2mm 2mm 0mm, clip]{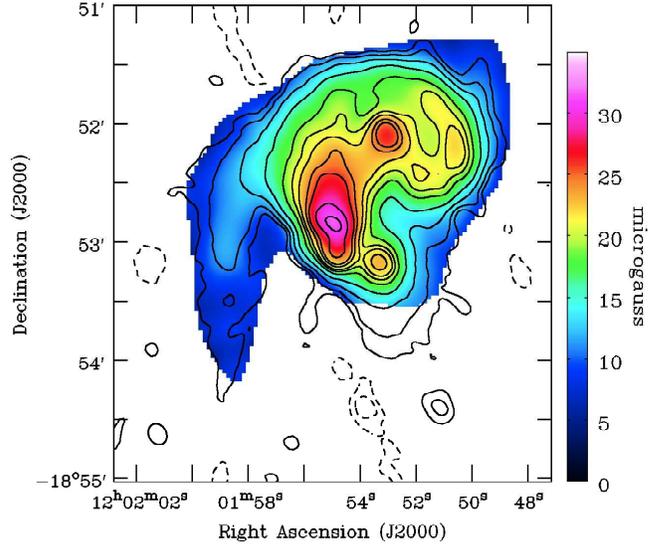}}}\\
\end{tabular}
\caption{Total magnetic field strength (in $\mu$G) estimated using the
non-thermal emission and by assuming energy equipartition between magnetic
fields and cosmic ray particles. Overlaid are the total intensity contours
same as Figure~\ref{totI}.} 
\label{totmag}
\end{figure}

The total magnetic field strength ($B_{\rm tot}$) was computed assuming
equipartition of energy between magnetic field strength and cosmic ray
particles using the revised equipartition formula by \citet{beckkrause05}.
Figure~\ref{totmag} shows the magnetic field strength map overlaid with the
total intensity contours.  We used the non-thermal emission at 2.8 GHz and the
non-thermal spectral index map to compute the field strength. In addition, we
assumed a synchrotron path length $l=2$ kpc, corrected for inclination and the
ratio of number densities of relativistic protons to electrons ($K_0$) of 100.
The average magnetic field strength in NGC 4038/9 is found to be $\sim
20~\mu$G, significantly stronger than that of normal star-forming galaxies
having typical magnetic field strength of $\sim 10~\mu$G \citep{basu13, eck15}.
Our estimated magnetic field strength is consistent with that obtained in
\citet{chyzy04}. Note that our assumption of a constant $l$ throughout the
Antennae galaxies is likely not realistic and the estimated magnetic field can
be scaled by $[2 \times 10^{-2} (K_0 + 1)/l_{\rm kpc}]^{1/(\ant + 3)}$ due to
the assumption of $K_0 =100$ and $l= 2$ kpc.  A factor of 2 difference in the
path length in different parts of the galaxies would change the field strength
estimate by less than 20 per cent.

Within the merging disc, the magnetic field is strong: from $\sim 10~\mu$G in
the periphery to $\sim25~\mu$G.  The magnetic field strength is strongest,
$\sim35~\mu$G, in the dark dusty cloud complex.  This strong magnetic field is
likely produced by the fluctuation dynamo \citep{kandu98} driven by high star
formation activity. We detect low surface brightness synchrotron emission from
the tidal tail with an equipartition field strength of $\sim 6-10~\mu$G.
However, because of significant energy loss of the CREs in the tidal tail (see
\textsection3.1.1), the energy spectral index between CREs and cosmic ray
proton differs. The energy spectral indices for cosmic ray protons and
electrons are assumed to be constant and the same in the revised equipartition
formula. We have therefore set $\ant$ as $-1$ in the tidal tail region and thus
the field strength are underestimated.

\subsubsection{Ordered magnetic field strength}  \label{bord}

\begin{table} 
\centering 
 \caption{The equipartition magnetic field strengths (in $\mu$G) of the
different magnetic field components at various region of the Antennae galaxies.
$B_{\rm tot}$ is the total strength, $B_{\rm turb}$ is the strength of the
turbulent field, $B_{\rm ord, \perp}$ is the ordered component in the plane of
the sky and $B_{\rm ord, \parallel}$ is the ordered component along the line of
sight. We have assumed a 3D isotropic turbulent field.}
\begin{tabular}{@{}lcccccc@{}} 
 \hline 
  Region & $B_{\rm tot}$ & $\frac{B_{\rm turb}}{B_{\rm ord, \perp}}$  & $\frac{B_{\rm turb}}{B_{\rm ord, \parallel}}$ & $B_{\rm turb}$  & $B_{\rm ord, \perp}$ & $B_{\rm ord, \parallel}$ \\ 
          &     ($\mu$G)  &   & &  ($\mu$G)&     ($\mu$G)&     ($\mu$G)\\ 
     (1)     &(2)       & (3) & (4) & (5) & (6) & (7) \\ 
\hline 
Southern tail    & 10  & 0.6  &  5.8--9.2 & 5  & 8.5 & $<1$  \\ 
Merging disc    & 20   & 7--12 & 4--5 & 19.5  & $<5$ & $<5$  \\ 
Northern arm    & 20   & $< 3.7$ & -- & 19  & $<6$ & --  \\ 
Star-forming    & 25--35  & -- & --  & --  & -- & --  \\ 
\hline 
\end{tabular}
\label{btable} 
\end{table}

The internal degree of polarization ($\Pi_{\rm int}$) is related to the ratio
of the 3D isotropic turbulent field ($B_{\rm turb}$) and the ordered magnetic
field in the plane of the sky ($B_{\rm ord,\perp}$).  Under the assumption of
equipartition, $\Pi_{\rm int}$ is related to 
$q=B_{\rm turb}/B_{\rm ord, \perp}$ by \citep{sokol98}
\begin{equation}
\Pi_{\rm int} = \Pi_{\rm max} \frac{\left(1 + \frac{7}{3} q^2\right)}{\left(1 + 3 q^2 + \frac{10}{9} q^4\right)}.
\end{equation}
We note that, the above relation assumes the ordered magnetic fields lies
entirely in the plane of the disc.  Thus, for a realistic scenario of a
non-zero ordered magnetic field component perpendicular to the disc, especially
for the merging galaxies, $B_{\rm ord, \perp}$ would be overestimated.
Furthermore, since, $B_{\rm tot}^2 = B_{\rm ord}^2 + B_{\rm turb}^2 = B_{\rm
turb}^2 \left[1 + q^{-2} + (B_{\rm ord, \parallel}/B_{\rm turb})^2 \right]$,
the above assumption does not affect the estimation of $B_{\rm turb}$ using $q$
if the turbulent field is isotropic, but it affects the regular field strength
estimation. This is because what we actually measure is $1/q_{\rm obs}^2 =
1/q^2 + (B_{\rm ord, \parallel}/B_{\rm turb})^2$.

At the frequency of our observations, the fractional polarization is strongly
affected by $\lambda-$dependent depolarization, therefore we are unable to
produce a map of $B_{\rm ord, \perp}$. We use the fitted values for $\Pi_{\rm
int}$ from \textsection\ref{depoltailmerger} to estimate $B_{\rm ord, \perp}$.
We further use the ratio $\sigmaRM/\langle {\rm RM} \rangle$ to estimate
$B_{\rm ord, \parallel}/B_{\rm turb}$ (see \textsection\ref{orderedtail}).  In
the tidal tail, the average $B_{\rm tot}$ is $\sim10~\mu$G, $B_{\rm turb}$ is
$\sim5~\mu$G and the $B_{\rm \perp, ord}$ is $\sim8.5~\mu$G. In the merging
disc, the total field strength is found to be $\sim20~\mu$G. This gives a
turbulent field strength of $19.5~\mu$G for the fitted $\Pi_{\rm int}$ of
$\sim0.04$ and the upper limit on the $B_{\rm ord, \perp}$ of $2.7~\mu$G. The
magnetic field strength estimates in different regions of the Antennae are
summarized in Table~\ref{btable}.

\section{Discussion}

\subsection{Regular field of $\sim20$ kpc in the tidal tail} \label{orderedtail}

We detect highly polarized emission along the southern tidal tail of the
Antennae with intrinsic polarization fraction close to the theoretical maximum
(see \textsection\ref{depoltailmerger}). The magnetic field orientations in the
plane of the sky corrected for Faraday rotation are well-aligned along the tail
(see Figure~\ref{polimap}). Such a high degree of polarization can originate
from a combination of anisotropic turbulent magnetic fields by compressing an
initially random field, and regular magnetic fields in the plane of the sky
\citep{beck16}. However, anisotropic fields dominating over the regular field
will not contribute to Faraday depth \citep{jaffe10}.  From our data, we find
the Milky Way foreground-corrected Faraday depth along the tidal tail varies
smoothly with positive sign throughout (see Figure~\ref{rmmap}) indicating the
line-of-sight ordered field points toward the observer.  Anisotropic random
fields alone cannot give rise to a smooth large-scale variation of Faraday
depth measured over several beams. We therefore conclude that the magnetic
field in the plane of the sky along the tidal tail is regular (or coherent) and
maintains its direction over $\sim20$ kpc. To our knowledge, this is the
largest coherent magnetic field structure observed in galaxies. We discuss in
detail the properties of the magneto-ionic medium in the tidal tail.

\subsubsection{Turbulent cell size}

The standard deviation of the Faraday depth within the 3D beam ($\sigmaRM$) in
the tidal tail region is found to be $131$ rad m$^{-2}$ (see
\textsection\ref{depoltailmerger}). This dispersion of the RM is caused by
fluctuations of the field strength along the line of sight.  The $\sigmaRM$
depends on the turbulent magnetic field along the line of sight and the
properties of the magneto-ionic medium as
\begin{equation}
\sigma_{\rm RM} = 0.81 \left(\frac{\langle n_{\rm e}\rangle}{\rm cm^{-3}}\right) \left(\frac{B_{\rm turb, \parallel}}{\rm \mu G}\right) \left(\frac{L_{\rm ion} d}{f}\right)^{1/2}.
\label{sigmaRM}
\end{equation}
Here, $\langle n_{\rm e}\rangle$ is the average thermal electron density along
the line of sight, $B_{\rm turb, \parallel}$ is the strength of the random
magnetic field along the line of sight, $L_{\rm ion}$ is the path length
through the ionized medium (in pc), $d$ is the size of the turbulent cells (in
pc) and $f$ is the volume filling factor of electrons along the line of sight.
The RM dispersion in the plane of the sky ($\sigma_{\rm RM, sky}$) is related
to the $\sigma_{\rm RM}$ as \citep{fletc11}
\begin{equation}
\sigma_{\rm RM, sky} \simeq N^{-1/2} \sigma_{\rm RM},
\end{equation}
where, $N\approx(D/d)^2$ is the number of turbulent cells for the projected
beam area in the sky of diameter $D$, for which $\sigma_{\rm RM, sky}$ is
measured.  Thus, we can compute the diameter of typical turbulent cell size $d$
using the observed $\sigma_{\rm RM, sky}$, $\sigma_{\rm RM}$, and the beam size
$D\approx 1400$ pc.  $d$ is found to be $\sim230$ pc, significantly larger than
the typical turbulent cell size of $\sim50$ pc observed in the discs of
galaxies \citep{fletc11, mao15, haver08}.  Hence, assuming the line of sight
extent of the tidal tail to be the same as the thickness in the plane of sky,
i.e.  $L_{\rm ion}=1.1$ kpc, the field along the line of sight must be regular.
Therefore, the observed dispersion of the Faraday depth must be caused by
systematic variations in the plane of the sky. The Faraday depth indeed varies
smoothly in the tidal tail (see Figure~\ref{rmmap}).

\subsubsection{Regular field strengths}

The mean RM depends on the regular component of the magnetic field along the
line of sight ($B_{\rm reg, \parallel}$) as
\begin{equation}
\langle {\rm RM}\rangle = 0.81 \langle n_{\rm e} \rangle B_{\rm reg, \parallel} L_{\rm ion \label{eqrm}}.
\end{equation}
Thus, the ratio of $\sigma_{\rm RM}$ to $\langle \rm RM \rangle$ can give
us the estimate of $B_{\rm turb, \parallel}/B_{\rm reg, \parallel}$ and is
given by
\begin{equation}
\frac{\sigmaRM}{|\langle \rm RM \rangle|} = \frac{B_{\rm turb, \parallel}}{|B_{\rm reg, \parallel}|}\left(\frac{d}{f L_{\rm ion}} \right)^{1/2}. 
\end{equation}
From our estimated values of $\sigmaRM$, $\langle \rm RM\rangle$ and $d$, and
assumed value for $L_{\rm ion}$, we find $B_{\rm turb, \parallel}/B_{\rm reg,
\parallel}\sim 13 f^{1/2}$.  Compared to the star-forming disc, the ionized
medium in the tidal tail is more diffuse.  Therefore, assuming typical $f$ in
the range $0.2 - 0.5$ for a diffuse medium, $B_{\rm turb, \parallel}/B_{\rm
reg, \parallel}$ lies in the range $5.8-9.2$.  Assuming $B_{\rm turb}=5~\mu$G
is isotropic (see \textsection\ref{bord}), we estimate $B_{\rm reg, \parallel}$
to be $\lesssim 1~\mu$G. Thus, the field strength in the plane of the sky
($B_{\rm reg, \perp} \sim8.5~\mu$G) is significantly larger than $B_{\rm reg,
\parallel}$, perhaps caused due to stretching and twisting of the remnant
spiral field in the disc of the galaxies during the tidal interaction. 

Here, we explore the degree of stretching required to amplify an initial
regular field in the progenitor galaxy to the observed field strengths in the
tidal tail. For this, we consider the scenario that an initially cylindrical
spiral-shaped regular field generated in the progenitor galaxies by dynamo
action has been stretched by the tidal interaction.  Assuming the field is
stretched keeping the cross-sectional area constant, then the ratio $B/\rho l$
is conserved if the magnetic flux is frozen in the tube \citep[see
e.g.,][]{longa11}. Here, $\rho$ is the density of the gas and $l$ is the length
of the cylinder. From observations of H{\sc i} gas \citep{hibba01}, the density
in the tidal tail is about a factor of 3 lower than that in the remnant spiral
arm.  Therefore, a regular field of 8.5 $\mu$G in the tidal tail requires
stretching by a factor of $\sim4-9$ for an initial regular field of
$\sim3-6~\mu$G in strength (see \textsection\ref{n-arm}).  This implies, the
dynamo generated initial field was regular over $\sim2-5$ kpc, which is a
typical length-scale observed in nearby spiral galaxies \citep{beck13book}. 
Thus, tidal stretching of field lines can also amplify large-scale regular
fields within the dynamical time-scale of the merger event, i.e., few 100 Myr.
A full treatment of the 3D magnetic field structure is beyond the scope of this
paper and will be discussed in a forthcoming paper (A.  Basu et al. 2016 in
preparation).

\subsubsection{Thermal electron densities}

Using Equation~\ref{eqrm}, and the upper limit on $B_{\rm \parallel, reg}$, we
constrain $\langle n_{\rm e}\rangle$ to be $\gtrsim0.02$ cm$^{-3}$ in the tidal
tail.  The column density of the H{\sc i} in the tidal tail is found to be
$\sim6\times10^{20}$ cm$^{-2}$ having thickness $\sim 4.5$ kpc \citep{hibba01}
which corresponds to $\langle n_{\rm H}\rangle \sim 0.04$ cm$^{-3}$.  Thus,
from our constraint on the $\langle n_{\rm e} \rangle$, the ionization fraction
in the tidal tail is $\gtrsim30$ per cent.  As pointed out in the study by
\citet{hibba05}, the UV emission from the tidal tail predominantly arises from
old stars, and there is little evidence of on-going star formation.  This is
insufficient to sustain the ionic medium and therefore the ionization of the
tidal tail is maintained by the inter-galactic radiation field.

\subsection{Turbulent fields in the merging disc}

The polarized emission from the merging disc of the galaxies has low fractional
polarization with a median value of only $\sim0.016$ at 2.8 GHz and the
$\lambda-$dependent depolarization is best described by an external dispersion
screen (see \textsection3.3).  From the fitted value of $\sigmaRM\approx100$
rad m$^{-2}$ and observed $|\langle \rm RM \rangle|\approx40$ rad m$^{-2}$, we
find $B_{\rm turb, \parallel}/B_{\rm ord, \parallel} \approx 16 f^{1/2}$,
assuming path length $L_{\rm ion}=2$ kpc and a typical turbulent cell size of
$\sim50$ pc observed in discs of galaxies. Hence, $B_{\rm turb,
\parallel}/B_{\rm ord, \parallel}$ ranges from 4--5 for $f$ in the range
0.05--0.1.  The measured low $\Pi_{\rm int}$ of 0.04 is likely caused by a
turbulent magnetic field enhanced due to the merger event.  For the estimated
$\Pi_{\rm int}$ of 0.04, we find $B_{\rm turb, \perp}/B_{\rm ord, \perp}$ in
the range 7--12, i.e., the turbulent field strength significantly
dominates over the ordered field both in the plane of the sky and along the
line of sight. For isotropic turbulence, $B_{\rm turb}$ is found to be
$\sim19.5~\mu$G. Using this, we constrain the strength of the regular fields
along the line of sight $B_{\rm ord, \perp}$ to be $\lesssim5~\mu$G, and in the
plane of the sky, $B_{\rm ord, \parallel} \approx 1.2f^{-1/2} \lesssim 5~\mu$G
for $f\gtrsim0.05$.

\subsection{Ordered fields in the northern arm}\label{n-arm}

\begin{figure}
\begin{centering}
\begin{tabular}{c}
{\mbox{\includegraphics[height=9cm]{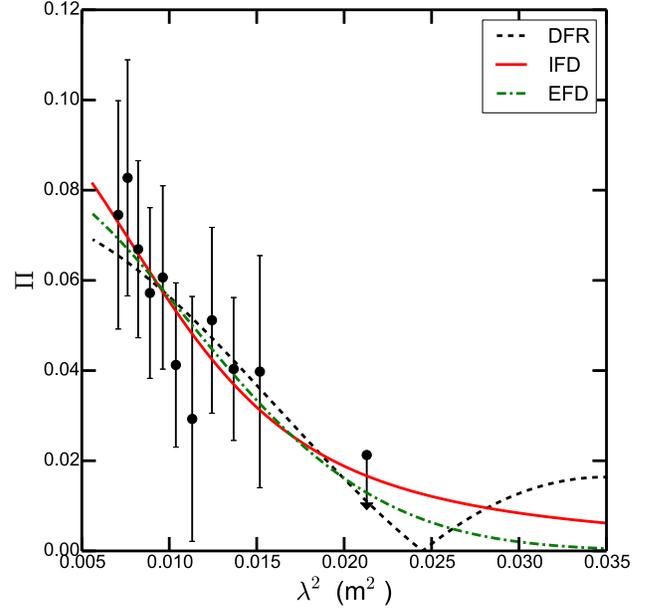}}}\\
\end{tabular}
\end{centering}
\caption{Variation of the fractional polarization ($\Pi$) with $\lambda^2$
around the northern spiral arm region. The point with arrow is the 2$\sigma$
upper limit. The lines are the best fit models of different types.}
\label{banddepolnorth}
\end{figure}

The polarized emission from the relic spiral arm of NGC 4038 in the north is
weak with median $\Pi\sim0.08$ at 2.8 GHz. The $B_{\rm \perp, reg}$ is
observed to follow a spiral pattern (see Figure~\ref{composite}). The polarized
emission is depolarized in the optical spiral arm hosting sites of high star
formation and the detected polarized emission is offset toward the outer parts.
It is difficult to reliably measure the fractional polarization as the total
intensity quickly falls off to the background noise level in that region. The
mean Faraday depth of this feature is $\sim+9$ rad m$^{-2}$ and has a
comparatively large dispersion of $\sim20$ rad m$^{-2}$ in the plane of the sky
varying between $-30$ to $+40$ rad m$^{-2}$ (see Figure~\ref{rmmap}).  We
observe the Faraday depth to frequently change sign smoothly over a few
synthesized beam, indicative of a less coherent regular component of the
magnetic field in the plane of the sky than that in the tidal tail.

In Figure~\ref{banddepolnorth}, we show the variation of $\Pi$ as a function of
$\lambda^2$ in the northern arm region. Due to a limited $\lambda^2$ coverage
of our data, it is not possible to select a best fit model from our data
($\chi^2=2.2, 1.6$ and $1.8$ for the DFR, IFD and EFD models, respectively).
All the depolarization mechanisms indicate a $\Pi_{\rm int}$ between 0.08--0.1
and if we assume isotropic turbulent fields, $B_{\rm turb}$ is estimated to be
$\sim19~\mu$G. Since, we do not have a estimate of $B_{\rm turb, \parallel}
/B_{\rm ord, \parallel}$ due to lack of unambiguous fit to Faraday
depolarization, we estimate an upper limit on the strength of $B_{\rm ord,
\perp}$ as $6~\mu$G.

\subsection{Extreme Faraday depolarization in southern star-forming regions}

The southern part of the Antennae system around the dark cloud region shows
strong depolarization, such that the polarized emission remains undetectable in
our observations below $\sim3.6$ GHz, at 4.8 GHz \citep{chyzy04} and the
fractional polarization at 8.44 GHz is extremely low \citep[$\sim1$
per cent;][]{chyzy04}. Although, such an effect could be caused by
$\lambda-$independent beam depolarization due to the random component of the
magnetic field, here we assess the possibility of extreme nature of Faraday
depolarization.

In \textsection\ref{sfreg} we showed, based on the estimated thermal emission,
we expect $|\phi|$ $\sim10^4$ rad m$^{-2}$. Such high $\phi$ values are
unlikely to cause bandwidth depolarization especially above $\sim3$ GHz within
a 8 MHz channel.  Assuming that the $\lambda-$dependent depolarization arises
due to the same region being Faraday rotating and synchrotron emitting but with
a regular magnetic field along the line of sight, i.e., depolarization due to
differential Faraday rotation (DFR), the $\Pi(\lambda)$ is given by
\begin{equation}
\Pi(\lambda) = \Pi_{\rm int} \frac{\sin |\phi \lambda^2|}{|\phi \lambda^2|}.
\end{equation}
In this case, $\phi$ is related to the RM as $\phi = (1/2)\textrm{RM}$.  Using
this we find, for the estimated $\phi\sim10^4$ rad m$^{-2}$, the expected
fractional polarization at 8.44 GHz with a bandwidth of 50 MHz \citep[same as
the observations of][]{chyzy04} lies between $1-7$ per cent with occasional
null values depending on the $|\phi|$. The variation of the fractional
polarization at 8.44 GHz along this region was observed to be spatially smooth
(1--2 per cent) and hence it is unlikely that an ionic medium with only a
uniform magnetic field gives rise to such high depolarization.

If the magneto-ionic medium is turbulent, driven by high star-formation
activity and merger, then the internal and external Faraday dispersion models,
give $\sigma_{\rm RM} \sim 10^3$ rad m$^{-2}$.  Using Equation~\ref{sigmaRM} we
infer the turbulent cells to be 10--50 pc in size for a turbulent field
strength of $\sim20~\mu$G and the estimated $\langle n_{\rm e}\rangle \approx
3~\rm cm^{-3}$. The cell size is typical of what is observed in the spiral arms
of normal star-forming galaxies \citep{fletc11, mao15}.  Hence, it is difficult
to disentangle whether the star-forming regions in the southern part are beam
depolarized due to a turbulent magnetic field or Faraday depolarized due to
extreme properties of the magneto-ionic media.  To distinguish between these
broadband properties require Stokes $Q$ and $U$ fitting of higher resolution
(A- and B-array) and higher frequency (4--8 GHz) data. We have acquired data
between 4 and 8 GHz using the VLA in the DnC, CnB and BnA array configurations
and the data will be analyzed.

\subsection{Implications on the buildup of galactic magnetic fields}

{\it Implications on ISM of galaxies}: As a late stage merger, the Antennae is
a classic example of a system at the peak of the cosmic star formation history.
We, however, note that the early merging galaxies are believed to be different
in their interstellar medium ISM properties as compared to the present day
mergers.  The present day merging systems, such as the Antennae galaxies, are
predominantly between well-settled, dynamically cold discs with comparatively
lower star formation rates as compared to early mergers which can be between
more gas-rich, turbulent, and compact systems \citep{forst09, willi11,
stott16}.  Based on the bolometric far infrared luminosity, the Antennae pair
is classified as a luminous infrared galaxy (LIRG). LIRGs contribute
significantly to the comoving star formation density beyond redshift of 1
\citep{magne09}.  Therefore, a detail understanding of the Antennae is
essential to understand cosmic evolution of ISM in galaxies. Merger induced
active star formation and turbulence \citep{veill02} is an essential ingredient
in the evolution of hot interstellar gas through stellar feedback
\citep{metz04}. It is crucial in the rapid amplification of magnetic fields via
turbulent-dynamo mechanisms. Our study demonstrates the magnetic field strength
in the Antennae is comparatively higher than that in isolated galaxies and is
dominated significantly by turbulent fields within the merging bodies. 

The turbulent magnetic field---and its coupling with the ISM energy densities,
especially the kinetic energy of the turbulent gas---is important in the origin
and maintenance of the radio--far infrared correlation at higher redshifts
\citep{schle13}. The stronger field strength in merging galaxies
($\sim20~\mu$G) compensate for the inverse-Compton losses due to the cosmic
microwave background at higher redshifts and helps in maintaining the
radio--far infrared correlation \citep{basu15b}.

{\it Comparison to the pan-Magellanic field}: In the immediate neighbourhood of
the Milky Way, the Magellanic bridge connecting the Large and Small Magellanic
clouds (LMC and SMC, respectively) could potentially be an example of a system
hosting regular magnetic fields of similar length-scale as detected in the
tidal tail of Antennae galaxies. Through studies of starlight polarization and
RMs inferred from background sources, it has been suggested that an aligned
``pan-Magellanic'' magnetic field possibly exists, connecting LMC and SMC
\citep{mao08, wayte90}.  \citet{deinz73} presented model of the H{\sc i} gas
connecting the two clouds that spans $\sim20$ kpc, which supports the existence
of pan-Magellanic magnetic field. Although, similar to the tidal tail of the
Antennae galaxies, the Magellanic bridge is likely generated from a tidal event
\citep{besla12, baghe13}, its progenitors are very different from those of the
Antennae galaxies both in terms of morphologies and their initial field
strengths.  Detailed studies of magnetic field properties in LMC and SMC have
revealed low total field strengths of $\sim3~\mu$G \citep{mao08, mao12}.
Therefore, low progenitor magnetic field strengths along with significantly
less tidal pressure because of shallow gravitational potential of the LMC--SMC
system as compared to that of the Antennae galaxies, the strength of the
pan-Magellanic ordered field could be low. For example, the model by
\citet{deinz73}, predicts the strength of the pan-Magellanic ordered field to
be sub-microgauss to a few microgauss.

The pan-Magellanic ordered field could span about 20 degrees on the sky.
Therefore, direct detection of the regular field from linearly polarized
emission is difficult because of low surface brightness and systematic
contribution from the Milky Way in the foreground. To firmly establish the
existence of such a field, improved RM grid experiment with a large number of
background polarized sources is essential as suggested by \citet{mao08}.  A
systematic comparison of the pan-Magellanic magnetic field with that of the
tidal tail of the Antennae galaxies is not possible until we have firm
observations about the strength and the structure of this pan-Magellanic
magnetic field.

{\it Implications on magnetic field measurements at high redshifts}: Usually,
one studies magnetic field properties in high redshift intervening objects by
measuring the excess RM towards quasar absorption line systems: Mg{\sc ii},
damped Lyman-$\alpha$ (DLA), sub-DLA \citep[see e.g.,][]{oren95, berne08,
joshi13, farne14}. It has been suggested that the sub-DLAs can originate from
neutral gas that lies $\gtrsim20$ kpc from the host galaxies and the absorbing
gas is likely stripped via tidal interaction and/or ram pressure
\citep{semba01, muzah16}.  The tidal tail of the Antennae with $N_{\rm HI}$ in
the range $\sim10^{20}-6\times10^{20}$ cm$^{-2}$, would be a classic example of
a DLA or sub-DLA (depending on at what distance the from the host galaxies the
line-of-sight intersects), when observed as a Lyman-$\alpha$ absorber against a
quasar at high redshifts.  Our study shows that such systems can host large
scale regular magnetic fields and gives rise to rotation measure when observed
at suitable viewing angles. It is therefore important to take into account that
the inferred field could come from tidally stripped gas, and not only from the
disc.

{\it Implications to magnetization of the intergalactic medium}: Coherent
magnetic fields in the outskirts of merging galaxies, extending in to the
intergalactic medium, assists in propagation of cosmic ray particles along the
field lines \citep[see e.g.,][]{cesar80, ptusk06}. Thus, apart from starburst
driven galactic winds, galaxy mergers can also play an important role in
magnetizing the intergalactic medium and enriching it with cosmic ray
particles.

{\it Implication to evolution of large-scale fields in galaxies}: In a study of
cosmological evolution of large- and small-scale magnetic fields in galaxies,
\citet{arsha09} predicted that major merger events dissipate the $10^{-6}$ G
large-scale disc fields down to several $10^{-8}$ G.  Our study comprehensively
shows that, although the pre-merger regular magnetic fields in the galactic
discs are mostly disrupted by the merger and are dominated by turbulent fields,
they can assist in producing larger-scale coherent fields several microgauss in
strength through field stretching. The detection of a $\sim20$ kpc coherent
magnetic field in the tidal tail indicates that large-scale fields can be
preserved even in advanced merging systems. Moreover, large-scale fields do not
necessarily require Gyr timescales to develop as predicted by
magneto-hydrodynamic (MHD) simulations \citep{arsha09, hanas09}. A detailed MHD
simulation is essential to understand the nature of the coherent magnetic
fields in the tidal tails and its implications in developing large-scale
regular fields observed in isolated galaxies in the local Universe.

\section{Summary}

We have studied the magnetic field properties of the spiral galaxies NGC
4038/9, also known as the Antennae galaxies, undergoing late stage merger.  The
galaxies were observed between 2 and 4 GHz using the VLA in DnC and CnB arrays
and studied the polarization properties using the combined DnC+CnB array data.
We summarize our main findings in this section.

\begin{enumerate}[(i)]

\item We estimated the thermal emission from the galaxies using FUV emission as
a tracer. We find the mean $\fth$ to be $\sim25$ per cent, significantly higher
than what is found in normal star-forming galaxies. The galaxy-integrated
$\ant$ is found to be $-1.11\pm0.03$. The $\ant$ value can have systematic
error up to $\sim10$ per cent due to uncertainties in the thermal emission.

\item We detect the total intensity radio continuum emission from the gas-rich
southern tidal tail extending up to $\sim18$ kpc at $3\sigma$ level. This
region shows steep non-thermal spectrum with $\ant$ lying in the range $-1.2$
to $-1.6$, indicating that the tail is composed of an older population of CREs.

\item Employing the technique of RM synthesis, we detect polarized emission
from the southern tidal tail extending up to $\sim20$ kpc.  The Faraday depth
along the tidal tail varies smoothly preserving its sign throughout. The
$\lambda-$dependent Faraday depolarization of this region is best described by
an internal Faraday dispersion model with $\sigmaRM = 131\pm23$ \rad and an
internal fractional polarization of $0.62\pm0.18$. 

\item Our result suggests that the magnetic field along the tidal tail is
highly regular up to a size of $\sim20$ kpc, the largest known coherent field
structure on galactic scales. In this region, the regular field lies mostly in
the plane of the sky with $B_{\rm reg, \perp} \approx 8.5\mu$G and dominates
over the turbulent field ($B_{\rm turb} \approx 5~\mu$G).  The large-scale
field is likely generated by stretching of an initial disc field by a factor of
4--9 over the merger's dynamical time-scale of few 100 Myr.

\item We estimate the ionization fraction in the tidal tail to be $\gtrsim30$
per cent, although there is no indication for on-going star formation that
could maintain the ionized medium.  Inter-galactic radiation field is likely
the main contributor of ionizing photons.

\item The remnant spiral arm in the northern galaxy NGC 4038 has spiral-shaped
regular magnetic field structures and it is displaced outward by $\sim1.6$ kpc
with respect to that of the optical and total intensity radio-continuum
emission. Here, the Faraday depth varies between $-30$ and $+40$ \rad and
frequently reverses sign along the arm, indicating the magnetic field is less
ordered than that in the tail.  The magnetic field strength in this region is
dominated by the turbulent field ($\sim19~\mu$G).

\item In the merging disc, the $\lambda-$dependent Faraday depolarization is
best described by an external Faraday dispersion screen with strong turbulent
fields of strength $\sim19.5~\mu$G.  Faraday depth spectra in this region have
complex structures with multiple and/or broad Faraday depth components.  The
ordered component of magnetic fields, $B_{\rm ord, \perp}$ and $B_{\rm ord,
\parallel}$ is estimated to be $<2.7$ and $<4~\mu$G, respectively.

\item The magnetic field strength is strongest in the star-forming regions
reaching values up to $\sim35~\mu$G. The polarized emission from the
star-forming regions remain undetected at the sensitivity of our observations.
This is perhaps caused due to turbulence driven by merger induced intense star
formation.

\end{enumerate}

\section*{Acknowledgments}

We thank Dr. Rainer Beck for critical review of the manuscript and useful
discussions that improved the presentation of the paper. We thank Prof. Sne{\v
z}ana Stanimirovi{\'c} for the help during carrying out of the VLA
observations.  Ancor Damas and Maja Kierdorf are acknowledged for carefully
reading the manuscript and helpful comments.  The VLA is operated by the
National Radio Astronomy Observatory (NRAO). The NRAO is a facility of the
National Science Foundation operated under cooperative agreement by Associated
Universities, Inc.  Based on observations made with the NASA/ESA Hubble Space
Telescope,  and obtained from the Hubble Legacy Archive, which is a
collaboration between the Space Telescope Science Institute (STScI/NASA), the
Space Telescope European Coordinating Facility (ST-ECF/ESA) and the Canadian
Astronomy Data Centre (CADC/NRC/CSA).  Some of the data presented in this paper
were obtained from the Mikulski Archive for Space Telescopes (MAST). STScI is
operated by the Association of Universities for Research in Astronomy, Inc.,
under NASA contract NAS5-26555.

\bibliographystyle{mn2e}

\bibliography{abasu_etal_mnr.bbl}

\begin{thebibliography}{77}
\expandafter\ifx\csname natexlab\endcsname\relax\def\natexlab#1{#1}\fi

\bibitem[{Ade {et~al.}(2015)Ade, Aghanim, Arnaud, Arroja, Ashdown, Aumont,
  Baccigalupi, Ballardini, Banday, \& et al.}]{ade15}
Ade P., Aghanim N., Arnaud M., Arroja F., Ashdown M., Aumont J., Baccigalupi
  C., Ballardini M., Banday A., et al., 2015, arXiv:1502.01594

\bibitem[{{Arshakian} {et~al.}(2009){Arshakian}, {Beck}, {Krause}, \&
  {Sokoloff}}]{arsha09}
{Arshakian} T.~G., {Beck} R., {Krause} M., {Sokoloff} D., 2009, \aap, 494, 21

\bibitem[{{Bagheri} {et~al.}(2013){Bagheri}, {Cioni}, \&
  {Napiwotzki}}]{baghe13}
{Bagheri} G., {Cioni} M.-R.~L., {Napiwotzki} R., 2013, \aap, 551, A78

\bibitem[{Basu {et~al.}(2012)Basu, Mitra, Wadadekar, \&
  Ishwara-Chandra}]{basu12}
Basu A., Mitra D., Wadadekar Y., Ishwara-Chandra C.~H., 2012, \mnras, 419, 1136

\bibitem[{Basu \& Roy(2013)}]{basu13}
Basu A., Roy S., 2013, \mnras, 433, 1675

\bibitem[{{Basu} {et~al.}(2015){Basu}, {Wadadekar}, {Beelen}, {Singh},
  {Archana}, {Sirothia}, \& {Ishwara-Chandra}}]{basu15b}
{Basu} A., {Wadadekar} Y., {Beelen} A., {Singh} V., {Archana} K.~N., {Sirothia}
  S., {Ishwara-Chandra} C.~H., 2015, \apj, 803, 51

\bibitem[{Beck(2007)}]{beck07}
Beck R., 2007, \aap, 470, 539

\bibitem[{{Beck}(2016)}]{beck16}
{Beck} R., 2016, \aapr, 24, 4

\bibitem[{Beck \& Krause(2005)}]{beckkrause05}
Beck R., Krause M., 2005, Astronomische Nachrichten, 326, 414

\bibitem[{Beck \& Wielebinski(2013)}]{beck13book}
Beck R., Wielebinski R., 2013, {Magnetic Fields in Galaxies}, Oswalt T.,
  Gilmore G., eds., p. 641

\bibitem[{Bell {et~al.}(2013)Bell, Oppermann, Crai, \& En{\ss}lin}]{bell13}
Bell M., Oppermann N., Crai A., En{\ss}lin T., 2013, \aap, 551, L7

\bibitem[{Bernet {et~al.}(2008)Bernet, Miniati, Lilly, Kronberg, \&
  Dessauges-Zavadsky}]{berne08}
Bernet M., Miniati F., Lilly S., Kronberg P., Dessauges-Zavadsky M., 2008,
  \nat, 454, 302

\bibitem[{{Besla} {et~al.}(2012){Besla}, {Kallivayalil}, {Hernquist}, {van der
  Marel}, {Cox}, \& {Kere{\v s}}}]{besla12}
{Besla} G., {Kallivayalil} N., {Hernquist} L., {van der Marel} R.~P., {Cox}
  T.~J., {Kere{\v s}} D., 2012, \mnras, 421, 2109

\bibitem[{{Brentjens} \& {de Bruyn}(2005)}]{brent05}
{Brentjens} M.~A., {de Bruyn} A.~G., 2005, \aap, 441, 1217

\bibitem[{Brindle {et~al.}(1991)Brindle, Hough, Bailey, Axon, \&
  Sparks}]{brind92}
Brindle C., Hough J., Bailey J., Axon D., Sparks W., 1991, \mnras, 252, 288

\bibitem[{{Cesarsky}(1980)}]{cesar80}
{Cesarsky} C.~J., 1980, \araa, 18, 289

\bibitem[{Chamandy {et~al.}(2013)Chamandy, Subramanian, \& Shukurov}]{chama13}
Chamandy L., Subramanian K., Shukurov A., 2013, \mnras, 428, 3569

\bibitem[{Chy\.zy \& Beck(2004)}]{chyzy04}
Chy\.zy K., Beck R., 2004, \aap, 417, 541

\bibitem[{Condon(1992)}]{condo92}
Condon J., 1992, \araa, 30, 575

\bibitem[{{Deinzer} \& {Schmidt}(1973)}]{deinz73}
{Deinzer} W., {Schmidt} T., 1973, \aap, 27, 85

\bibitem[{Drzazga {et~al.}(2011)Drzazga, Chy\.zy, Jurusik, \&
  Wi{\'{o}}rkiewicz}]{drzaz11}
Drzazga R., Chy\.zy K., Jurusik W., Wi{\'{o}}rkiewicz K., 2011, \aap, 533, A22

\bibitem[{{Ehle} \& {Beck}(1993)}]{ehle93}
{Ehle} M., {Beck} R., 1993, \aap, 273, 45

\bibitem[{Farnes {et~al.}(2014)Farnes, O'Sullivan, Corrigan, \&
  Gaensler}]{farne14}
Farnes J., O'Sullivan S., Corrigan M., Gaensler B., 2014, \apj, 795, 63

\bibitem[{{Federrath} {et~al.}(2011){Federrath}, {Chabrier}, {Schober},
  {Banerjee}, {Klessen}, \& {Schleicher}}]{feder11}
{Federrath} C., {Chabrier} G., {Schober} J., {Banerjee} R., {Klessen} R.~S.,
  {Schleicher} D.~R.~G., 2011, Physical Review Letters, 107, 114504

\bibitem[{Fletcher {et~al.}(2011)Fletcher, Beck, Shukurov, Berkhuijsen, \&
  Horellou}]{fletc11}
Fletcher A., Beck R., Shukurov A., Berkhuijsen E., Horellou C., 2011, \mnras,
  412, 2396

\bibitem[{{F{\"o}rster Schreiber} {et~al.}(2009){F{\"o}rster Schreiber},
  {Genzel}, {Bouch{\'e}}, {Cresci}, {Davies}, {Buschkamp}, {Shapiro},
  {Tacconi}, {Hicks}, {Genel}, {Shapley}, {Erb}, {Steidel}, {Lutz},
  {Eisenhauer}, {Gillessen}, {Sternberg}, {Renzini}, {Cimatti}, {Daddi},
  {Kurk}, {Lilly}, {Kong}, {Lehnert}, {Nesvadba}, {Verma}, {McCracken},
  {Arimoto}, {Mignoli}, \& {Onodera}}]{forst09}
{F{\"o}rster Schreiber} N.~M., {Genzel} R., {Bouch{\'e}} N., {Cresci} G.,
  {Davies} R., {Buschkamp} P., {Shapiro} K., {Tacconi} L.~J., et al.,
  \apj, 706, 1364

\bibitem[{Gaensler {et~al.}(2005)Gaensler, Haverkorn, Staveley-Smith, Dickey,
  McClure-Griffiths, Dickel, \& Wolleben}]{gaens05}
Gaensler B., Haverkorn M., Staveley-Smith L., Dickey J., McClure-Griffiths N.,
  Dickel J., Wolleben M., 2005, Science, 307, 1610

\bibitem[{{Hanasz} {et~al.}(2009){Hanasz}, {Otmianowska-Mazur}, {Kowal}, \&
  {Lesch}}]{hanas09}
{Hanasz} M., {Otmianowska-Mazur} K., {Kowal} G., {Lesch} H., 2009, \aap, 498,
  335

\bibitem[{Hao {et~al.}(2011)Hao, Kennicutt, Johnson, Calzetti, Dale, \&
  Moustakas}]{hao11}
Hao C.-N., Kennicutt R., Johnson B., Calzetti D., Dale D., Moustakas J., 2011,
  \apj, 741, 124

\bibitem[{{Haverkorn} {et~al.}(2008){Haverkorn}, {Brown}, {Gaensler}, \&
  {McClure-Griffiths}}]{haver08}
{Haverkorn} M., {Brown} J.~C., {Gaensler} B.~M., {McClure-Griffiths} N.~M.,
  2008, \apj, 680, 362

\bibitem[{Hibbard {et~al.}(2001)Hibbard, van~der Hulst, Barnes, \&
  Rich}]{hibba01}
Hibbard J., van~der Hulst J., Barnes J., Rich R., 2001, \aj, 122, 2969

\bibitem[{{Hibbard} {et~al.}(2005){Hibbard}, {Bianchi}, {Thilker}, {Rich},
  {Schiminovich}, {Xu}, {Neff}, {Seibert}, {Lauger}, {Burgarella}, {Barlow},
  {Byun}, {Donas}, {Forster}, {Friedman}, {Heckman}, {Jelinsky}, {Lee},
  {Madore}, {Malina}, {Martin}, {Milliard}, {Morrissey}, {Siegmund}, {Small},
  {Szalay}, {Welsh}, \& {Wyder}}]{hibba05}
{Hibbard} J.~E., {Bianchi} L., {Thilker} D.~A., {Rich} R.~M., {Schiminovich}
  D., {Xu} C.~K., {Neff} S.~G., {Seibert} M., et al., 2005, \apjl, 619,
  L87

\bibitem[{Hummel \& Beck(1995)}]{humme95}
Hummel E., Beck R., 1995, \aap, 303, 691

\bibitem[{{Jaffe} {et~al.}(2010){Jaffe}, {Leahy}, {Banday}, {Leach}, {Lowe}, \&
  {Wilkinson}}]{jaffe10}
{Jaffe} T.~R., {Leahy} J.~P., {Banday} A.~J., {Leach} S.~M., {Lowe} S.~R.,
  {Wilkinson} A., 2010, \mnras, 401, 1013

\bibitem[{{Joshi} \& {Chand}(2013)}]{joshi13}
{Joshi} R., {Chand} H., 2013, \mnras, 434, 3566

\bibitem[{Karl {et~al.}(2010)Karl, Naab, Johansson, Kotarba, Boily, Renaud, \&
  Theis}]{karl10}
Karl S., Naab T., Johansson P., Kotarba H., Boily C., Renaud F., Theis C.,
  2010, \apjl, 715, L88

\bibitem[{Kennicutt \& Evans(2012)}]{kenni12}
Kennicutt R., Evans N., 2012, \araa, 50, 531

\bibitem[{Klaas {et~al.}(2010)Klaas, Nielbock, Haas, Krause, \&
  Schreiber}]{klaas10}
Klaas U., Nielbock M., Haas M., Krause O., Schreiber J., 2010, \aap, 518, L44

\bibitem[{Kotarba {et~al.}(2010)Kotarba, Karl, Naab, Johansson, Dolag, Lesch,
  \& Stasyszyn}]{kotar10}
Kotarba H., Karl S., Naab T., Johansson P., Dolag K., Lesch H., Stasyszyn F.,
  2010, \apj, 716, 1438

\bibitem[{Kulsrud \& Zweibel(2008)}]{kulsr08}
Kulsrud R.~M., Zweibel E.~G., 2008, Reports on Progress in Physics, 71, 046901

\bibitem[{Longair(2011)}]{longa11}
Longair M., 2011, {High energy astrophysics, 3rd ed}. Cambridge: Cambridge
  University Press

\bibitem[{Magnelli {et~al.}(2009)Magnelli, Elbaz, Chary, Dickinson, {Le
  Borgne}, Frayer, \& Willmer}]{magne09}
Magnelli B., Elbaz D., Chary R., Dickinson M., {Le Borgne} D., Frayer D.,
  Willmer C., 2009, \aap, 496, 57

\bibitem[{Mao {et~al.}(2008)Mao, Gaensler, Stanimirovi{\'{c}}, Haverkorn,
  McClure-Griffiths, Staveley-Smith, \& Dickey}]{mao08}
Mao S., Gaensler B., Stanimirovi{\'{c}} S., Haverkorn M., McClure-Griffiths N.,
  Staveley-Smith L., Dickey J., 2008, \apj, 688, 1029

\bibitem[{Mao {et~al.}(2015)Mao, Zweibel, Fletcher, Ott, \& Tabatabaei}]{mao15}
Mao S., Zweibel E., Fletcher A., Ott J., Tabatabaei F., 2015, \apj, 800, 92

\bibitem[{{Mao} {et~al.}(2012){Mao}, {McClure-Griffiths}, {Gaensler},
  {Haverkorn}, {Beck}, {McConnell}, {Wolleben}, {Stanimirovi{\'c}}, {Dickey},
  \& {Staveley-Smith}}]{mao12}
{Mao} S.~A., {McClure-Griffiths} N.~M., {Gaensler} B.~M., {Haverkorn} M.,
  {Beck} R., {McConnell} D., {Wolleben} M., {Stanimirovi{\'c}} S., {Dickey}
  J.~M., {Staveley-Smith} L., 2012, \apj, 759, 25

\bibitem[{{Metz} {et~al.}(2004){Metz}, {Cooper}, {Guerrero}, {Chu}, {Chen}, \&
  {Gruendl}}]{metz04}
{Metz} J.~M., {Cooper} R.~L., {Guerrero} M.~A., {Chu} Y.-H., {Chen} C.-H.~R.,
  {Gruendl} R.~A., 2004, \apj, 605, 725

\bibitem[{Mihos {et~al.}(1993)Mihos, Bothun, \& Richstone}]{mihos93}
Mihos J., Bothun G., Richstone D., 1993, \apj, 418, 82

\bibitem[{{Muzahid} {et~al.}(2016){Muzahid}, {Kacprzak}, {Charlton}, \&
  {Churchill}}]{muzah16}
{Muzahid} S., {Kacprzak} G.~G., {Charlton} J.~C., {Churchill} C.~W., 2016,
  \apj, 823, 66

\bibitem[{Niklas {et~al.}(1997)Niklas, Klein, \& Wielebinski}]{nikla97a}
Niklas S., Klein U., Wielebinski R., 1997, \aap, 322, 19

\bibitem[{Oppermann {et~al.}(2012)Oppermann, Junklewitz, Robbers, Bell,
  En{\ss}lin, Bonafede, Braun, Brown, Clarke, Feain, Gaensler, Hammond,
  Harvey-Smith, Heald, Johnston-Hollitt, Klein, Kronberg, Mao,
  McClure-Griffiths, O'Sullivan, Pratley, Robishaw, Roy, Schnitzeler,
  Sotomayor-Beltran, Stevens, Stil, Sunstrum, Tanna, Taylor, \& {Van
  Eck}}]{opper12}
Oppermann N., Junklewitz H., Robbers G., Bell M., En{\ss}lin T., Bonafede A.,
  Braun R., Brown J., et al., 2012, \aap, 542, A93

\bibitem[{{Oren} \& {Wolfe}(1995)}]{oren95}
{Oren} A.~L., {Wolfe} A.~M., 1995, \apj, 445, 624

\bibitem[{O'Sullivan {et~al.}(2012)O'Sullivan, Brown, Robishaw, Schnitzeler,
  McClure-Griffiths, Feain, Taylor, Gaensler, Landecker, Harvey-Smith, \&
  Carretti}]{sulli12}
O'Sullivan S., Brown S., Robishaw T., Schnitzeler D., McClure-Griffiths N.,
  Feain I., Taylor A., Gaensler B., et al.,
  2012, \mnras, 421, 3300

\bibitem[{{Pakmor} {et~al.}(2014){Pakmor}, {Marinacci}, \&
  {Springel}}]{pakmo14}
{Pakmor} R., {Marinacci} F., {Springel} V., 2014, \apjl, 783, L20

\bibitem[{Patton {et~al.}(2002)Patton, Pritchet, Carlberg, Marzke, Yee, Hall,
  Lin, Morris, Sawicki, Shepherd, \& Wirth}]{patto02}
Patton D., Pritchet C., Carlberg R., Marzke R., Yee H., Hall P., Lin H., Morris
  S., Sawicki M., Shepherd C., Wirth G., 2002, \apj, 565, 208

\bibitem[{Perley \& Butler(2013{\natexlab{a}})}]{perle13a}
Perley R., Butler B., 2013{\natexlab{a}}, \apjs, 204, 19

\bibitem[{Perley \& Butler(2013{\natexlab{b}})}]{perle13b}
---, 2013{\natexlab{b}}, \apjs, 206, 16

\bibitem[{Ptuskin(2006)}]{ptusk06}
Ptuskin V., 2006, Journal of Physics: Conference Series, 47, 113

\bibitem[{Rampazzo {et~al.}(2008)Rampazzo, Bonoli, \& Giro}]{rampa08}
Rampazzo R., Bonoli C., Giro E., 2008, Astronomische Nachrichten, 329, 855

\bibitem[{Rau \& Cornwell(2011)}]{rau11}
Rau U., Cornwell T., 2011, \aap, 532, A71

\bibitem[{Ruzmaikin {et~al.}(1988)Ruzmaikin, Sokolov, \& Shukurov}]{ruzma88}
Ruzmaikin A., Sokolov D., Shukurov A., eds., 1988, Astrophysics and Space
  Science Library, Vol. 133, {Magnetic fields of galaxies}

\bibitem[{Schleicher \& Beck(2013)}]{schle13}
Schleicher D., Beck R., 2013, \aap, 556, A142

\bibitem[{{Schober} {et~al.}(2013){Schober}, {Schleicher}, \&
  {Klessen}}]{schob13}
{Schober} J., {Schleicher} D.~R.~G., {Klessen} R.~S., 2013, \aap, 560, A87

\bibitem[{Schweizer {et~al.}(2008)Schweizer, Burns, Madore, Mager, Phillips,
  Freedman, Boldt, Contreras, Folatelli, Gonz{\'{a}}lez, Hamuy, Krzeminski,
  Morrell, Persson, Roth, \& Stritzinger}]{schwe08}
Schweizer F., Burns C., Madore B., Mager V., Phillips M., Freedman W., Boldt
  L., Contreras C., et al., 2008, \aj, 136, 1482

\bibitem[{{Sembach} {et~al.}(2001){Sembach}, {Howk}, {Savage}, \&
  {Shull}}]{semba01}
{Sembach} K.~R., {Howk} J.~C., {Savage} B.~D., {Shull} J.~M., 2001, \aj, 121,
  992

\bibitem[{Sokoloff {et~al.}(1998)Sokoloff, Bykov, Shukurov, Berkhuijsen, Beck,
  \& Poezd}]{sokol98}
Sokoloff D., Bykov A., Shukurov A., Berkhuijsen E., Beck R., Poezd A., 1998,
  \mnras, 299, 189

\bibitem[{{Stott} {et~al.}(2016){Stott}, {Swinbank}, {Johnson}, {Tiley},
  {Magdis}, {Bower}, {Bunker}, {Bureau}, {Harrison}, {Jarvis}, {Sharples},
  {Smail}, {Sobral}, {Best}, \& {Cirasuolo}}]{stott16}
{Stott} J.~P., {Swinbank} A.~M., {Johnson} H.~L., {Tiley} A., {Magdis} G.,
  {Bower} R., {Bunker} A.~J., {Bureau} M., et al., 2016,
  \mnras, 457, 1888

\bibitem[{{Subramanian}(1998)}]{kandu98}
{Subramanian} K., 1998, \mnras, 294, 718

\bibitem[{{Subramanian}(1999)}]{kandu99}
---, 1999, Physical Review Letters, 83, 2957

\bibitem[{Tabatabaei {et~al.}(2007)Tabatabaei, Beck, Kr{\"{u}}gel, Krause,
  Berkhuijsen, Gordon, \& Menten}]{tabat07b}
Tabatabaei F., Beck R., Kr{\"{u}}gel E., Krause M., Berkhuijsen E., Gordon K.,
  Menten K., 2007, \aap, 475, 133

\bibitem[{{Van Eck} {et~al.}(2015){Van Eck}, Brown, Shukurov, \&
  Fletcher}]{eck15}
{Van Eck} C., Brown J., Shukurov A., Fletcher A., 2015, \apj, 799, 35

\bibitem[{Veilleux {et~al.}(2002)Veilleux, Kim, \& Sanders}]{veill02}
Veilleux S., Kim D.-C., Sanders D., 2002, \apjs, 143, 315

\bibitem[{{Wardle} \& {Kronberg}(1974)}]{wardl74}
{Wardle} J.~F.~C., {Kronberg} P.~P., 1974, \apj, 194, 249

\bibitem[{{Wayte}(1990)}]{wayte90}
{Wayte} S.~R., 1990, \apj, 355, 473

\bibitem[{{Whitmore} {et~al.}(1999){Whitmore}, {Zhang}, {Leitherer}, {Fall},
  {Schweizer}, \& {Miller}}]{whitm99}
{Whitmore} B.~C., {Zhang} Q., {Leitherer} C., {Fall} S.~M., {Schweizer} F.,
  {Miller} B.~W., 1999, \aj, 118, 1551

\bibitem[{Williams {et~al.}(2011)Williams, Quadri, \& Franx}]{willi11}
Williams R.~J., Quadri R.~F., Franx M., 2011, \apj, 738, L25

\bibitem[{Zhang {et~al.}(2010)Zhang, Gao, \& Kong}]{zhang10}
Zhang H.-X., Gao Y., Kong X., 2010, \mnras, 401, 1839

\bibitem[{Zhang {et~al.}(2001)Zhang, Fall, \& Whitmore}]{zhang01}
Zhang Q., Fall S., Whitmore B., 2001, \apj, 561, 727

\end{thebibliography}

\bsp

\appendix

\section{Thermal emission separation}\label{thermal}

The radio continuum emission is mainly a combination of non-thermal synchrotron
and thermal free--free emission.  Although, the overall contribution of thermal
emission to the total radio emission at 1.4 GHz is $\sim10$ per cent for normal
star-forming galaxies \citep{nikla97a, tabat07b, basu12}, the thermal
fraction\footnote{Thermal fraction at a frequency $\nu$ is defined as, $f_{\rm
th, \nu}=S_{\rm \nu, th}/S_{\rm \nu, tot}$. Here, $S_{\rm \nu, th}$ is the flux
density of the thermal component of the total emission $S_{\rm \nu, tot}$.  We
express $\fth$ in per cent.} ($\fth$) could be much higher---more than 30 per
cent in star-forming regions.  Detailed studies of the star formation history
in the Antennae have revealed intense starbursts in localized regions,
especially in the overlapping region and the western galaxy NGC 4038
\citep{zhang10, klaas10}.  Such regions are bright in H$\alpha$,
coincident with the peaks in the total radio continuum emission (see
Figure~\ref{totI}) due to enhanced thermal emission.  We estimated the thermal
emission on a pixel-by-pixel basis using star formation rate (SFR) as its
tracer following \citet{condo92}. The thermal emission ($S_{\rm\nu, th}$) at a
radio frequency $\nu$ is related to $\textrm {SFR}$ as
\begin{equation}
\left(\frac{S_{\rm\nu, th}}{\rm Jy}\right) \approx 4.6 D_{\rm Mpc}^{-2} \nu_{\rm GHz}^{-0.1} \left( \frac{\textrm {SFR}}{\rm M_\odot yr^{-1}}\right).
\end{equation}
Here, $D_{\rm Mpc}$ is the distance to the galaxy in Mpc. We used the {\it
GALEX} far-ultraviolet (FUV; $\lambda\approx1520 \AA$)
image\footnote{Downloaded from the {\it MAST} website.} to estimate SFR using
the calibration given in \citet{kenni12}. We corrected the FUV emission for 
dust extinction using the observed FUV--NUV colour ($(FUV-NUV)_{\rm obs}$;
\citealt{hao11})
\begin{equation}
A_{\rm FUV} = (3.83\pm0.48)\times [(FUV-NUV)_{\rm obs} - (0.022\pm0.024)].
\end{equation}
The large error in the attenuation ($A_{\rm FUV}$) calibration gives rise
to up to $\sim20$ per cent error in extinction correction for the range of
$(FUV-NUV)_{\rm obs}$ in the Antennae galaxies. The highest extinction is
observed around the central regions of the two galaxies and along the spiral
arms. In those regions, the estimated SFR, and hence the thermal emission, can
be uncertain by at most 30 per cent. We estimate the total $\textrm{SFR}$ for
the Antennae galaxies to be $\sim10~\rm M_\odot~yr^{-1}$ which is in agreement
with total $\textrm{SFR}$ of $13~\rm M_\odot~yr^{-1}$ estimated by
\citet{klaas10} using total infrared emission within 30 per cent.\footnote{Note
that the $\textrm{SFR}$ in \citet{klaas10} was estimated to be $22~\rm
M_\odot~yr^{-1}$. The difference arises due their assumed distance of 28.4 Mpc.
However, the assumed distance does not affect the surface brightness of the
thermal emission.}

\begin{figure}
\begin{centering}
\begin{tabular}{c}
\includegraphics[width=8cm, trim=6mm 2mm 2mm 0mm, clip]{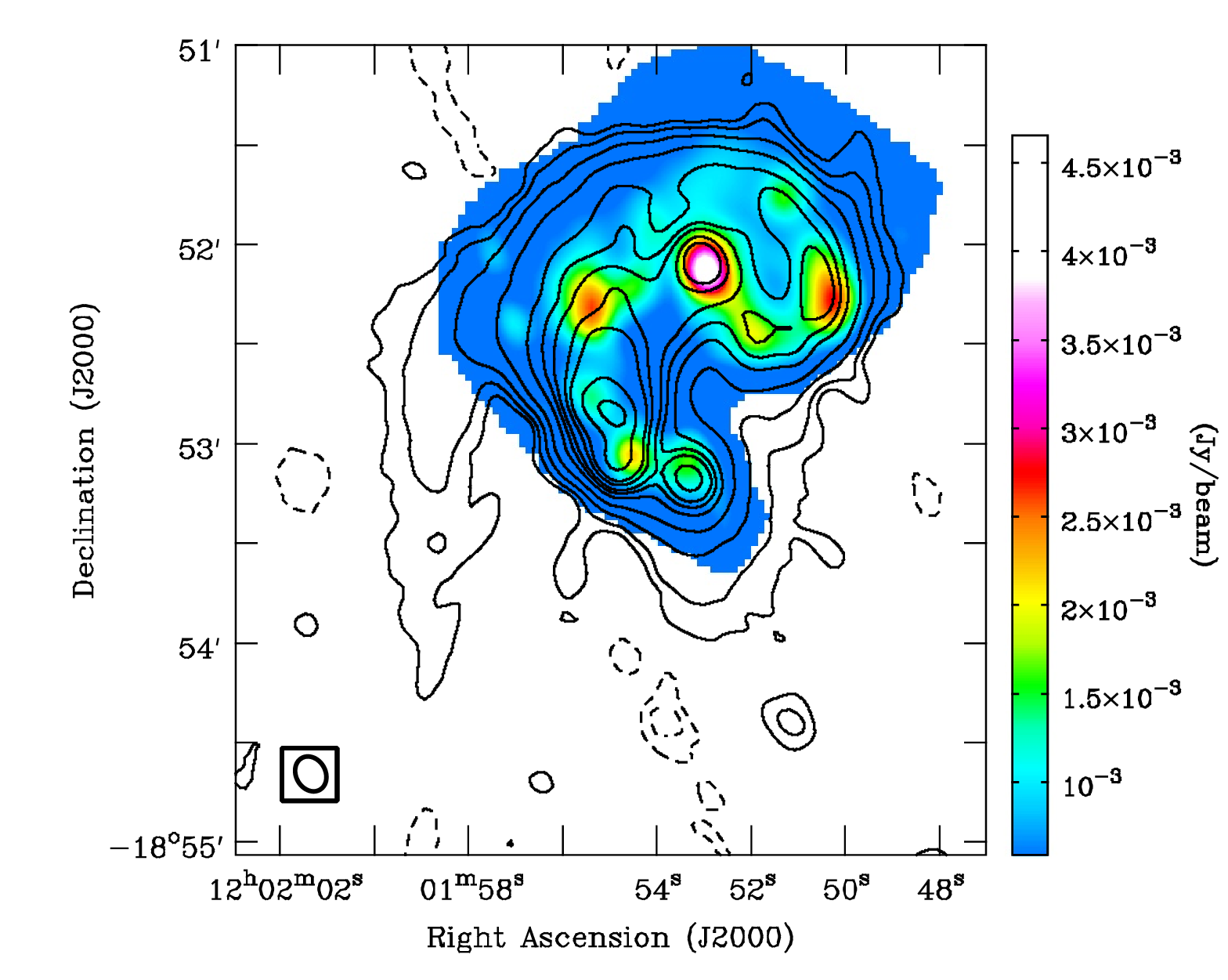}\\
\end{tabular}
\end{centering}
\caption{Thermal emission map of the Antennae galaxies estimated using
extinction corrected FUV emission. The overlaid total intensity contours are
same as Figure~\ref{totI}.}
\label{sth}
\end{figure}

Our estimated total $\textrm{SFR}$ of 10 $\rm M_\odot~yr^{-1}$ corresponds to a
total thermal emission of $86$ mJy, hence mean $f_{\rm th}\sim25$ per cent at
2.8 GHz. In Figure~\ref{sth}, we show the thermal emission map of the
Antennae galaxies. The estimated $\fth$ is in good agreement with 27 per cent
at 2.8 GHz estimated by interpolating $\fth\approx50$ per cent at 10.45 GHz for
the flux density scale assumed in \citet{chyzy04}. The mean $\fth$ is
significantly higher than what have been observed for normal star-forming
galaxies.  The western galaxy NGC 4038 have comparatively higher $\fth$ of
$\sim30$ per cent at the centre and $\sim24$ per cent in its remnant spiral
arm. These regions host intense star formation likely induced by the merger
\citep{zhang10}. In the dark cloud region, radio emission is significantly
dominated by non-thermal emission where $\fth\lesssim8$ per cent. This is also
the region where $\fth$ is lowest. 

We note that our thermal emission could be overestimated because of FUV
emission was used as a tracer of star formation. The thermal emission is best
traced by H$\alpha$ emission originating from ionization by young ($\lesssim10$
Myr) and massive ($\gtrsim10~\rm M_\odot$) stars. While, the FUV emission could
also originate from an older population (10--100 Myr) of lower mass stars
\citep{kenni12}.  The contribution of the older stellar population to the FUV
emission therefore overestimates the recent SFR and the thermal emission can be
overestimated. We could not use the H$\alpha$ emission for estimating the
thermal emission because the only publicly-available continuum subtracted
H$\alpha$ map does not cover the entire Antennae galaxies \citep[see
e.g.,][]{whitm99}.

\begin{figure*}
\begin{centering}
\begin{tabular}{cc}
\includegraphics[width=8cm, trim=5mm 10mm 2mm 0mm, clip]{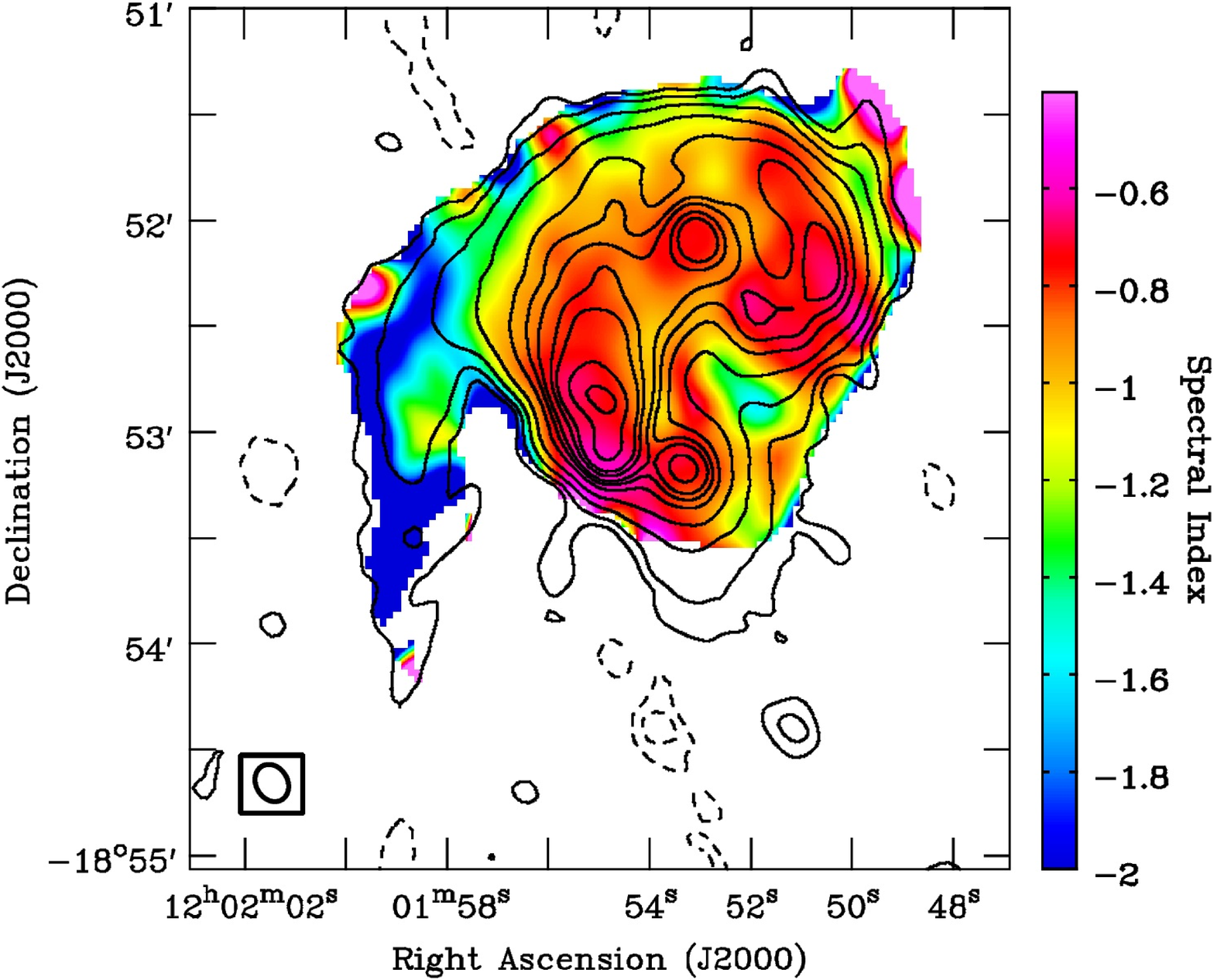}&
\includegraphics[width=8cm, trim=2mm 10mm 2mm 0mm, clip]{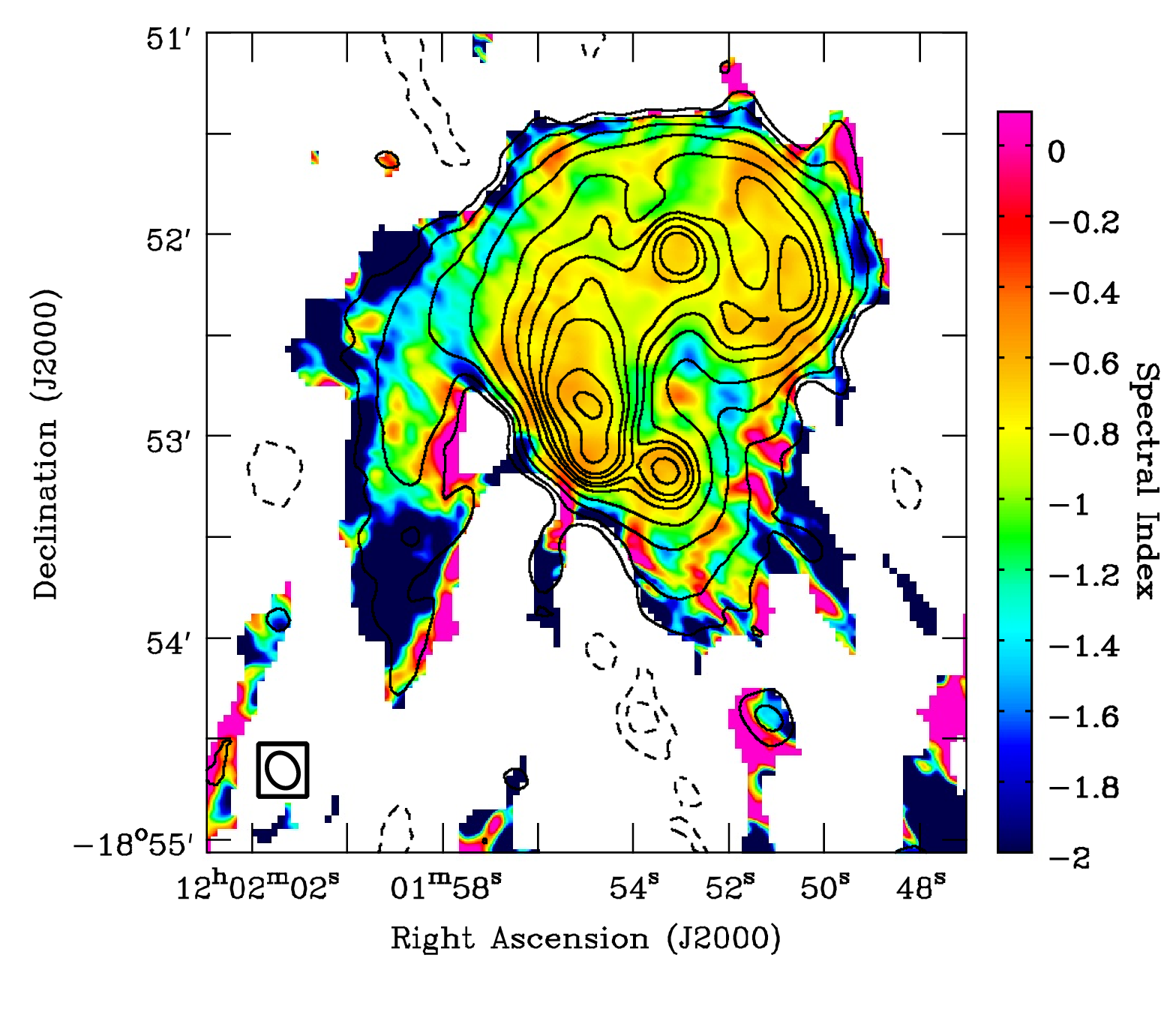}\\
\end{tabular}
\end{centering}
\caption{{\it Left-hand panel:} In-band spectral index map between 2 and 3.6
GHz at an angular resolution $15\times 12$ arcsec$^2$ produced by power-law
fitting of total intensity images of each spectral window on a pixel-by-pixel
basis.  {\it Right-hand panel:} Spectral index map produced using the {\sc
clean} task in {\sc casa} with {\it nterms=2} at an angular resolution
$11\times 9$ arcsec$^2$.  The overlaid total intensity contours are same as
Figure~\ref{totI}. Note that, at low signal-to-noise ratio regions, especially
in the outer parts, the spectral index computed by {\sc casa} gives rise to
unphysically flat or inverted spectrum, unlike in the left-hand panel.}
\label{totspind}
\end{figure*}

\section{Spectral index distribution}\label{spind_distr}

The total intensity maps of each spectral window were used to compute the
in-band spectral index ($\alpha$) between 2 and 3.6 GHz of the Antennae
galaxies.  The spectral index map for NGC 4038/9 was computed by fitting a
power law of the form, $\log S_\nu = \beta + \alpha \log \nu$, to each pixel of
the total intensity maps across the 11 spectral window images.  Here, $S_\nu$
is the total flux density at a frequency $\nu$ and $\beta$ is the normalization
constant at $\nu=1$ GHz.  We convolved the total intensity images of all the 11
spectral windows to a common resolution of $15\times12$ arcsec$^2$ (the
resolution at the lowest frequency). All the maps were then aligned to a common
coordinate system to do the fitting. The left-hand panel of
Figure~\ref{totspind} shows the spectral index map.  The galaxy-integrated
spectral index between 2--3.6 GHz, estimated by fitting the integrated total
intensities shown in Figure~\ref{integ}, is found to be $-0.85\pm0.02$.  The
spectral index shows large variations between $-0.6$ in the dark cloud complex
and $<-1.2$ in the tidal tail.  The typical error on the fitted values of
spectral index in high signal-to-noise ratio (SNR; $\gtrsim10$) pixels is
$\lesssim5$ per cent.  In low SNR ($\lesssim5$) pixels, especially in the outer
parts and tidal tail, the spectral index error is up to $\sim25$ per cent. The
variation of the spectral index is in good agreement with \citet{chyzy04}. 

The task {\sc clean} with {\it nterms=2} in {\sc casa} can also produce a
spectral index map through modelling the sky brightness as a Taylor-series
expansion per frequency channel \citep[see][]{rau11}.  In right-hand panel of
Figure~\ref{totspind} we show the spectral index map computed using {\sc casa}.
Note that, the colour-scale of the two maps are different in order to represent
the full range of values.  In regions with high SNR ($\gtrsim6$) in inner parts
of the Antennae, both the methods are in excellent agreement with each other.
But, in regions with $\textrm{SNR}\lesssim5$, especially in the outer parts of
the extended emission, the spectrum computed by {\sc casa} is either flat or
inverted which is unrealistic. Our fitting does not produce $\alpha$ with
values $\gtrsim-0.3$.  

The differences in $\alpha$ in the outer parts is not critical for our
analysis, but we prefer to use the pixel-by-pixel fitting method as it enables
us to compute the non-thermal spectral index map.  The non-thermal spectral
index, $\ant$, was computed after subtracting the thermal emission map scaled
to the frequencies of each spectral window.  A similar pixel-by-pixel fitting
was done using the non-thermal emission map of each spectral window. The green
dash-dot line in Figure~\ref{integ} represents the thermal emission with a
spectral index of $-0.1$. The non-thermal emission at each spectral window
after subtracting the thermal emission from the total intensity are shown as
the red squares. The galaxy-integrated $\ant$ estimated by fitting the total
non-thermal emission at each spectral window is found to be $-1.11\pm0.03$ and
is shown as the red dashed line in Figure~\ref{integ}. However, the estimated
$\ant$, overall, can have systematic error up to $\sim10$ per cent due to
uncertainty in the estimated thermal emission. In spatially resolved case, the
uncertainty in the thermal emission affects $\ant$ up to 10 per cent in bright
regions where $\fth$ is high. The errors in $\fth$ do not affect $\ant$ in
regions of low thermal fraction, i.e., in the outer parts of the Antennae and
in the tidal tail. Our estimated value of the $\ant$ is significantly steeper
than the value ($-0.8$) assumed by \citet{chyzy04} for performing a crude
thermal emission separation. We note that, the estimated $\ant$ is a lower
limit because the thermal emission could be overestimated by the FUV emission
(see Appendix~\ref{thermal}).

\label{lastpage}

\end{document}